\documentclass[aps,prd,superscriptaddress,nofootinbib,amsmath,amsfonts,preprintnumbers,groupedaddress,showpacs,
10pt,english]{revtex4-1}
\usepackage{amsmath}
\usepackage{amssymb}
\usepackage{babel}
\usepackage{wrapfig}
\usepackage{cancel}
\usepackage{bigints}

\newcommand{\e}{\mathrm{e}}
\usepackage{relsize,exscale}
\makeatletter

\usepackage{array,multirow,graphicx}
\usepackage{dcolumn}
\usepackage{newlfont}
\usepackage{bm}
\usepackage[colorlinks,citecolor=blue,urlcolor=blue,linkcolor=blue]{hyperref}
\usepackage[figtopcap]{subfigure}
\usepackage{color}
\usepackage{xcolor}

\def\mathcolor#1#{\@mathcolor{#1}}
\def\@mathcolor#1#2#3{%
 \protect\leavevmode
 \begingroup
 \color#1{#2}#3%
 \endgroup
}
\makeatother

\def\ba{\begin{align}}
\def\ea{\end{align}}

\def\be{\begin{align}}
\def\ee{\end{align}}

\allowdisplaybreaks[4]

\begin{document}

\tolerance=5000

\title{Isotropic compact stars in 4-dimensional Einstein-Gauss-Bonnet gravity \\
coupled with scalar field \\
-- Reconstruction of model ---}
\author{G.~G.~L.~Nashed}
\email{nashed@bue.edu.eg}
\affiliation {Centre for Theoretical Physics, The British University, P.O. Box
43, El Sherouk City, Cairo 11837, Egypt}
\author{Shin'ichi~Nojiri}
\email{nojiri@gravity.phys.nagoya-u.ac.jp}
\affiliation{Department of Physics, Nagoya University, Nagoya 464-8602,Japan \\\& \\
Kobayashi-Maskawa Institute for the Origin of Particles and the Universe,
Nagoya University, Nagoya 464-8602, Japan }
\date{}

\begin{abstract}
Recently, it has been supposed that the Einstein-Gauss- Bonnet theory coupled with scalar field (EGBS) maybe appropriately
admit physically viable models of celestial phenomena such that the scalar field effect is active in standard four dimensions.
We consider the spherically symmetric and static configuration of the compact star and explain the consequences
of the EGBS theory in the frame of stellar modeling.
In our formulation, for any given static profile of the energy density $\rho$ with the spherical symmetry and the arbitrary equation of state (EoS) of matter,
we can construct the model which reproduces the profile.
Because the profile of the energy density determines the mass $M$ and the radius $R_s$ of the compact star,
an arbitrary relation between the mass $M$ and the radius $R_s$ of the compact star can be realized by adjusting the potential
and the coefficient function of the Gauss-Bonnet term in the action of EGBS theory.
This could be regarded as a degeneracy between the EoS and the functions characterizing the model, which tells that only
the mass-radius relation is insufficient to constrain the model.
For example, we investigate a novel class of analytic spherically symmetric interior solutions by the polytropic EoS.
{ We discuss our model in detail and show} that it is in agreement with the necessary physical conditions required
for any realistic compact star approving that EGBS theory is consistent with observations.
\end{abstract}

\maketitle

\section{Introduction}\label{S1}

Although the fact that general relativity (GR) theory by Einstein is successful at present, that can forecast
and elucidate the increase of observational data.
Meanwhile, there are strong motivations to expect that it must be modified, due to its shortage in the quantization of gravity
and explaining the recent observational; puzzles in modern cosmology yielding to the study of amended theories of gravity.

The Lovelock gravitational theories \cite{1971JMP....12..498L} are of special attraction since they are Lagrangian-based theories
that could give conserved covariant field equations which do not include the derivatives higher than the second degree.
In this regard, Lovelock's theories are the physical extensions of GR.
The Gauss-Bonnet (GB) theory is considered the first physical non-trivial expansion of Einstein's GR.
This theory is meaningful if its space-time is greater than 4-dimensional in which the GB invariant
\begin{align}
\mathcal{G} = R^2-4R_{\alpha \beta}R^{\alpha \beta}+R_{\alpha \beta \rho \sigma}R^{\alpha \beta \rho \sigma}\, ,
\label{eq:GB}
\end{align}
can create a rich phenomenology.
Through the use of Chern's theorem \cite{Chern1945OnTC}, it can be shown that in 4 dimensions, the GB expression is a non-dynamical term
because the GB invariant becomes a total derivative.
To make the GB expression a dynamical one in 4 dimensions, we must invoke a novel scalar field with a canonical kinetic term coupling
to GB term \cite{Sotiriou:2013qea, Sotiriou:2014pfa,Kanti:1995vq, Kleihaus:2011tg,Doneva:2017bvd,Silva:2017uqg, Antoniou:2017acq,Dima:2020yac,
Herdeiro:2020wei, Berti:2020kgk} as stimulated, for example, by low-energy effective actions stem out of string theory, e.g.,
the Einstein-dilaton-GB models \cite{1985PhLB..156..315Z,Kanti:1995vq,Kanti:1997br,Cunha:2016wzk}.
Actually, because of Lovelock's theorem, all amended gravitational theories in 4 dimensions in principle will have extra degrees of freedom,
which can be considered as new basic fields.

The exact solutions of the gravitational system supply scientific society with a simple test of space-time and evaluation of observable forecasts.
Nevertheless, amended gravitational theories with new basic field(s) usually provide equations of motion with high intractability
so that the evaluations become analytically out of the question.
To face such an issue, one has to be coerced to apply either the perturbation theory, which is not well-qualified
in the strong gravitational field or to defy numerical methods \cite{Sullivan:2019vyi}.
However, the field equations of GR coupled with a matter having conformal invariance since it possesses
the constant Ricci scalar curvature on-shell, limiting the space-times and permitting analytic solutions to be easily derived.
An example of such a theory that has conformal invariance and yields simple analytic solutions is the electro-vacuum,
whose Reissner-Nordstr\"om (Kerr-Newman) solution was the first-ever discovered static (spinning) black hole (BH) with a matter source.
Another model is the gravitational theory coupled with a conformally scalar field, in which the matter action obeys the conformal invariance and has the form,
\begin{align}
S_\xi = \int d^4x \sqrt{-g} \left(\frac{1}{6} R \xi^2 +\left( \nabla \xi\right)^2\right)\, .
\label{eq:confcoupledaction}
\end{align}
where $R$ is the Ricci scalar and $\xi$ is the scalar field.
The field equations of the above action give a solution with no-hair theorems (see e.g., Ref.~\cite{Herdeiro:2015waa} for a review)
and { the static Bocharova-Bronnikov-Melnikov-Bekenstein BH \cite{Bocharova:1970skc,Bekenstein:1975ts,Bekenstein:1974sf} has been much debated}.
Gravitational theory with a conformal scalar field and its solutions have been discussed throughout recent years because of its compelling properties
(see e.g. Refs.~\cite{Martinez:2002ru,Martinez:2005di,Anabalon:2009qt,Padilla:2013jza,Fernandes:2021dsb,deHaro:2006ymc,Dotti:2007cp,
Gunzig:2000yj,Oliva:2011np,Cisterna:2021xxq,Caceres:2020myr} and references therein).

As we discussed above, in 4 dimensions, the GB term is topological and does not yield any dynamical effect.
Nevertheless, when the GB term is non-minimally coupled with any other field like a scalar field $\xi$, the output dynamics are non-trivial.
Many cosmological proposals have been presented in recent literature \cite{Brax:2003fv,Nojiri:2005jg,Nojiri:2005am,Cognola:2006eg,Nojiri:2010oco,
Cognola:2009jx,Capozziello:2008gu,Maharaj:2015gsd,bamba2008future,Sadeghi:2009pu,Guo:2010jr,Satoh:2010ep,Nozari:2013wua,Lahiri:2016qih,lahiri2016anisotropic,Nashed:2018cth,
Mathew:2016anx,Nozari:2015jca,Motaharfar:2016dqt,Carter:2005fu,DeLaurentis:2015fea,vandeBruck:2016xvt,Granda:2014zea,Granda:2011kx,Nojiri:2005vv,
Hikmawan:2015rze,Kanti:1998jd,Easther:1996yd,Rizos:1993rt,Starobinsky:1980te,brandenberger1989superstrings,tseytlin1992elements,Nashed:2016tbj,
Mukhanov:1991zn,Brandenberger:1993ef,Barrow:1993hp,Kobayashi:2005ch,Brassel:2018ybl,Damour:1994zq,Maeda:2005ci,Dehghani:2009si,Nashed:2011fg,
Angelantonj:1994dv,Kaloper:1995tu,Gasperini:1996in,Rey:1996ad,Rey:1996ka,Easther:1995ba,Santillan:2017nik,Bose:1997qv,KalyanaRama:1996ar,Nashed:2021sji,
KalyanaRama:1996im,KalyanaRama:1997xt,Brustein:1997ny,Brustein:1997cv}
and references therein.
{ In the frame of astrophysics, however, as far as we know, the GB theory with a non-minimal coupling of a scalar field via potential and coefficient function}
has not been tackled although there are some frontier works as in \cite{Silva:2017uqg}.
It is the aim of the present study to derive exact spherically symmetric interior solutions of this theory and discussed
the obtained physical consequences.
By using our formulation, we can construct a model which reproduces any given profile of the energy density $\rho$ for arbitrary EoS of matter.
The mass $M$ and the radius $R_s$ of the compact star are determined by the profile of the energy density and therefore
we can obtain an arbitrary relation between the mass $M$ and the radius $R_s$ of the compact star by adjusting the scalar potential
and the coefficient function of the Gauss-Bonnet term in the action of EGBS, which could be a kind of degeneracy between the EoS
and the functions characterizing the model
Therefore { we find that only the mass-radius relation is not sufficient to constrain the model.}

The arrangement of the present study is as follows:
In Section~\ref{S1}, we give the cornerstone of { the Einstein-Gauss-Bonnet gravity coupled with a scalar field (EGBS)}.
In Section~\ref{S2}, we apply the field equation of { the EGBS} theory to a spherically symmetric space-time and derive the full system of the differential equation.
Here we show that we can construct a model which reproduces any given profile of the energy density $\rho$ for arbitrary EoS of matter.
Also in Section~\ref{S2}, we give the form of a polytropic equation of state (EoS) as an example and a form of one of the metric potentials as an input
and then derived all the unknown functions
{ including the profile of the scalar field, the coefficient function, the potential of the scalar field, and the form of another metric potential}.
Section~\ref{S3} states the physical conditions that must be satisfied for any real stellar configuration.
In Section~\ref{S4}, we discuss the physical properties analytically and graphically showing that { the solutions} have realistic physical properties.
In Section~\ref{S5}, we discuss the issue of stability by using the adiabatic index { and show that our model satisfies the adiabatic index,
that is, the value of the index is greater than $4/3$,} which is the condition of stability.
The final Section is reserved for the conclusion and discussion of the present study.

\section{Gauss-Bonnet theory coupled with scalar through $f(\xi)$ }\label{S1}

Now we are going to consider { the Einstein-Gauss-Bonnet gravity coupled with a scalar field (EGBS)} in $N$ dimensions.
This theory takes the following amended action,
\begin{align}
\label{g2}
\mathcal{S}=\int d^N x \sqrt{-g}\left\{ \frac{1}{2\kappa^2}R
 - \frac{1}2 \partial_\mu \xi \partial^\mu \xi+V(\xi)+ f(\xi) \mathcal{G} \right\}+ S_\mathrm{M}\,,
\end{align}
where $\xi$ is the scalar field and $V$ is the potential which is a function of $\xi$, $f(\xi)$ is an arbitrary function of the scalar field, and $S_\mathrm{M}$
is the matter action, where we assume that matter is to couple minimally to the metric, i.e., we are working in the so-called Jordan frame.
In 4 dimensions, i.e., when $N=4$ the aforementioned action is physically non-trivial because the Gauss-Bonnet
invariant term $\mathcal{G}$ is coupled with the real scalar field $\xi$ through the coupling $f(\xi)$.
Because of this coupling, the Lagrangian is not a total derivative but contributes to the field equations of the system.

The variation of the action (\ref{g2}) w.r.t. the scalar field $\xi$ yields the following equation,
\begin{align}
\label{g3}
\nabla^2 \xi-V'(\xi)+ f'(\xi)\mathcal{G}=0\, .
\end{align}
The variation of the action (\ref{g2}) w.r.t. the metric $g_{\mu\nu}$ yields the following field equations,
\begin{align}
\label{GBeq}
T^{\mu \nu}=& \frac{1}{2\kappa^2}\left(- R^{\mu\nu} + \frac{1}2 g^{\mu\nu} R\right)+\frac{1}2 \partial^\mu \xi \partial^\nu \xi
 - \frac{1}{4}g^{\mu\nu} \partial_\rho \xi \partial^\rho \xi+ \frac{1}2 g^{\mu\nu}[f(\xi)G -V(\xi)]
+2 f(\xi) R R^{\mu\nu} \nonumber \\
& + 2 \nabla^\mu \nabla^\nu \left(f(\xi)R\right)- 2 g^{\mu\nu}\nabla^2\left(f(\xi)R\right)
+ 8f(\xi)R^\mu_{\ \rho} R^{\nu\rho}- 4 \nabla_\rho \nabla^\mu \left(f(\xi)R^{\nu\rho}\right)
 - 4 \nabla_\rho \nabla^\nu \left(f(\xi)R^{\mu\rho}\right) \nonumber \\
& + 4 \nabla^2 \left( f(\xi) R^{\mu\nu} \right)+ 4g^{\mu\nu} \nabla_{\rho} \nabla_\sigma \left(f(\xi) R^{\rho\sigma} \right)
- 2 f(\xi) R^{\mu\rho\sigma\tau}R^\nu_{\ \rho\sigma\tau}+ 4 \nabla_\rho \nabla_\sigma \left(f(\xi) R^{\mu\rho\sigma\nu}\right).
\end{align}
Through the use of the below Bianchi identities,
\begin{align}
\label{Bianchi}
\nabla^\rho R_{\rho\tau\mu\nu}=& \nabla_\mu R_{\nu\tau} - \nabla_\nu R_{\mu\tau} \, , \nonumber \\
\nabla^\rho R_{\rho\mu} =& \frac{1}2 \nabla_\mu R\, , \nonumber \\
\nabla_\rho \nabla_\sigma R^{\mu\rho\nu\sigma} =& \nabla^2 R^{\mu\nu} - {1 \over 2}\nabla^\mu \nabla^\nu R
+ R^{\mu\rho\nu\sigma} R_{\rho\sigma}- R^\mu_{\ \rho} R^{\nu\rho}\, , \nonumber \\
\nabla_\rho \nabla^\mu R^{\rho\nu} + \nabla_\rho \nabla^\nu R^{\rho\mu}
=& \frac{1}2 \left(\nabla^\mu \nabla^\nu R + \nabla^\nu \nabla^\mu R\right)
 - 2 R^{\mu\rho\nu\sigma} R_{\rho\sigma} + 2 R^\mu_{\ \rho} R^{\nu\rho}\, , \nonumber \\
\nabla_\rho \nabla_\sigma R^{\rho\sigma} =& \frac{1}2 \Box R \, ,
\end{align}
in Eq.~(\ref{GBeq}), we { obtain}
\begin{align}
\label{gb4b}
T^{\mu \nu}= & \frac{1}{2\kappa^2}\left(- R^{\mu\nu} + \frac{1}2 g^{\mu\nu} R\right)
+ \left(\frac{1}2 \partial^\mu \xi \partial^\nu \xi
 - \frac{1}{4}g^{\mu\nu} \partial_\rho \xi \partial^\rho \xi \right)
+ \frac{1}2 g^{\mu\nu} \left[ f(\xi) G-V(\xi) \right] \nonumber \\
& -2 f(\xi) R R^{\mu\nu} + 4f(\xi)R^\mu_{\ \rho} R^{\nu\rho}
 -2 f(\xi) R^{\mu\rho\sigma\tau}R^\nu_{\ \rho\sigma\tau}
 -4 f(\xi) R^{\mu\rho\sigma\nu}R_{\rho\sigma} \nonumber \\
& + 2 \left( \nabla^\mu \nabla^\nu f(\xi)\right)R
 - 2 g^{\mu\nu} \left( \nabla^2f(\xi)\right)R
 - 4 \left( \nabla_\rho \nabla^\mu f(\xi)\right)R^{\nu\rho}
 - 4 \left( \nabla_\rho \nabla^\nu f(\xi)\right)R^{\mu\rho} \nonumber \\
& + 4 \left( \nabla^2 f(\xi) \right)R^{\mu\nu}
+ 4g^{\mu\nu} \left( \nabla_{\rho} \nabla_\sigma f(\xi) \right) R^{\rho\sigma}
- 4 \left(\nabla_\rho \nabla_\sigma f(\xi) \right) R^{\mu\rho\nu\sigma}.
\end{align}
The field equations (\ref{g3}) and (\ref{gb4b}) are the full system of equations describing the theory under consideration.
In the 4 dimensional case i.e., $N=4$, Eq.~(\ref{gb4b}) yields to:
\begin{align}
\label{gb4bD4}
T^{\mu \nu}= & \frac{1}{2\kappa^2}\left(- R^{\mu\nu} + \frac{1}2 g^{\mu\nu} R\right)
+ \left(\frac{1}2 \partial^\mu \xi \partial^\nu \xi
 - \frac{1}{4}g^{\mu\nu} \partial_\rho \xi \partial^\rho \xi \right)
 - \frac{1}2 g^{\mu\nu}V(\xi) \nonumber \\
& + 2 \left( \nabla^\mu \nabla^\nu f(\xi)\right)R
 - 2 g^{\mu\nu} \left( \nabla^2f(\xi)\right)R
 - 4 \left( \nabla_\rho \nabla^\mu f(\xi)\right)R^{\nu\rho}
 - 4 \left( \nabla_\rho \nabla^\nu f(\xi)\right)R^{\mu\rho} \nonumber \\
& + 4 \left( \nabla^2 f(\xi) \right)R^{\mu\nu}
+ 4g^{\mu\nu} \left( \nabla_{\rho} \nabla_\sigma f(\xi) \right) R^{\rho\sigma}
- 4 \left(\nabla_\rho \nabla_\sigma f(\xi) \right) R^{\mu\rho\nu\sigma}.
\end{align}
In the present study, we assume { that the scalar field $\xi$ is a function of the radial coordinate $r$ and therefore the function $f(\xi)$ only depends on $r$, i.e.,
$f(r)\equiv f\left(\xi\left(r\right) \right)$, because we deal with static and spherically symmetric space-time,
\begin{align}
\label{met1}
ds^2 = -a(r)dt^2 +\frac{dr^2}{a_1(r)}+r^2 \left( d\theta^2 + \sin^2 \left(\theta\right)\right) d\phi^2\,.
\end{align}
}
In the following section, we { study the system of field equations (\ref{g3}) and (\ref{gb4bD4})}
and try to find the analytic form of the unknown functions when $T^{\mu \nu}\neq 0$.

\section{4 dimensional spherically symmetric interior solution in EGBS }\label{S2}

{ For the metric in (\ref{met1}),} the $(t,t)$-component of the field equation Eq.~(\ref{gb4bD4}) has the following form,
\begin{align}
\label{Eq2tt}
 -\rho =\frac{16a_1\left(1 -a_1\right) f'' + \left\{ 8\left( 1-3a_1 \right)f' +2r \right\} a'_1
+2a_1 +2V r^2 +r^2 \xi'^2 a_1 -2}{4r^2} \,,
\end{align}
the $(r,r)$-component is given by
\begin{align}
\label{Eq2rr}
p=\frac{ 2\left(4 \left(1 -3a_1 \right) f' +r \right)a_1a'+a\left[2a_1 -r^2 a_1\xi'^2 -2 +2V r^2 \right]}{4 r^2 a}\,,
\end{align}
and the $(\theta,\theta)$ and $(\phi,\phi)$-components are,
\begin{align}
\label{Eq2pp}
p=\frac{2a_1 a \left(r - 8 f' a_1 \right)a'' -16 f''a_1^2 a a' + a_1\left( 8 f' a_1- r \right) a'^2
+ \left\{ \left( r-24 f' a_1 \right)a'_1 +2 a_1 \right\} aa'
+2a^2 \left( a'_1 + r\left[ \xi'^2a_1 +2V \right] \right)}{8a^2r} \,.
\end{align}
The field equation of the scalar field (\ref{g3}) takes the following form
\begin{align}
\label{Eqphi2}
0=\frac{8a_1 a f' \left(a_1-1 \right) a'' +2 \xi'' a_1 \xi' a^2r^2 +4 a'f' \left[ a_1 a' \left(1-a_1 \right)
+ a'_1 a\left( 3a_1 - 1 \right) \right] +ra\left( \left[ \,a_1 a' r+a \left\{ 4 a_1
+a'_1 r \right\} \right]\xi'^2 -2a rV' \right) }{2 r^2 a^2 \xi'} \,.
\end{align}
Here $\rho$ is the energy density and $p$ is the pressure of matter, which we
assume to be a perfect fluid { and} satisfies an equation of state, $p=p\left(\rho\right)$.
The energy density $\rho$ and the pressure $p$ satisfy the following conservation law,
\begin{align}
\label{FRN2}
0 = \nabla^\mu T_{\mu r} =\frac{1}{2}\frac{a'}{a} \left( \rho + p \right) + \frac{dp}{dr} \, .
\end{align}
The conservation law is also derived from Eqs.~(\ref{Eq2tt}), (\ref{Eq2rr}), (\ref{Eq2pp}), and (\ref{Eqphi2}).
Here we have assumed $\rho$ and $p$ only depend on the radial coordinate $r$.
Other components of the conservation law are trivially satisfied.
If the equation of state $\rho=\rho(p)$ is given, Eq.~(\ref{FRN2}) can be integrated as
\begin{align}
\label{FRN3}
\frac{1}{2} \ln a = - \int^r dr \frac{\frac{dp}{dr}}{\rho + p}
= - \int^{p(r)}\frac{dp}{\rho(p) + p} \, .
\end{align}
Because Eq.~(\ref{FRN2}) and therefore (\ref{FRN3}) can be obtained from Eqs.~(\ref{Eq2tt}), (\ref{Eq2rr}), (\ref{Eq2pp}), and
(\ref{Eqphi2}), as long as we use (\ref{FRN3}), we forget one equation in Eqs.~(\ref{Eq2tt}), (\ref{Eq2rr}), (\ref{Eq2pp}), and (\ref{Eqphi2}).
In the following, we do not use Eq.~(\ref{Eqphi2}).
Inside the compact star, we can use Eq.~(\ref{FRN3}) but outside the star, we cannot use Eq.~(\ref{FRN3}).
Instead of using Eq.~(\ref{FRN3}), we may assume the profile of $a=a(r)$ so that $a(r)$ and $a'(r)$ are continuous
at the surface of the compact star.

By Eq.~(\ref{Eq2tt}) $+$ Eq.~(\ref{Eq2rr}), we obtain
\begin{align}
\label{V2}
V = -\rho + p
+\frac{8\,a_1a \left( a_1-1 \right) f'' -4\left\{\left( 1-3a_1 \right) f'+r \right\} \left( aa_1\right)' +2a-2aa_1}{2a r^2} \, .
\end{align}
On the other hand, Eq.~(\ref{Eq2tt}) $-$ Eq.~(\ref{Eq2rr}) gives,
\begin{align}
\label{xi2}
\xi' = \pm \left\{ \frac{2}{a_1} \left(\rho + p \right)
 + {\frac{8\,a a_1 \left(a_1 -1 \right)f'' - \left[ 4 \left( 1-3\,a_1 \right) f' +r \right] \left(a a'_1 -a_1 a' \right) }{a_1 r^2a }}
 \right\}^\frac{1}{2}\, .
\end{align}
Furthermore, Eq.~(\ref{Eq2tt}) $-$ Eq.~(\ref{Eq2pp}) gives,
\begin{align}
\label{f2}
0 =&16\, \left[ a_1 a' r-2\,a \left( a_1-1 \right) \right] a a_1 f'' -2ra a_1 \left(r -8\,f' a_1 \right)a''
+ra_1 \left(r -8\, f' a_1\right)a'^2 - \left[ \left( r-24\,f' a_1 \right)a'_1 +2\,a_1 \right] a ra'\nonumber \\
& +2\, \left\{ \left[8\left( 1-3\,a_1 \right) f' +r \right] a'_1 -2+2\,a_1 \right\} a^2
+ 8a^2 r^2 \left( \rho + p \right) \, ,
\end{align}
which can be regarded with the differential equation for $f'$ and therefore for $f$ if $a=a(r)$, $a_1=a_1(r)$, $\rho=\rho(r)$, and $p=p(r)$
are given and the solution is given by
\begin{align}
\label{f3}
f(r)=\, &-\int \left( \int {\frac{\left[a_1 a'^2 r^2 -2a_1 a'' a r^2-ra \left(a'_1 r+2\,a_1 \right)a' + {2\left\{a'_1 r-2+2a_1
+4 \left( \rho+p \right) r^2 \right\} a^2} \right] }
{2U(r) a_1 a \left( a_1a'r-2a \left(a_1-1 \right) \right) }}{dr}-16 c_1 \right) U dr +c_2
 \, , \nonumber \\
U(r) \equiv\, & \e^{\int \frac{r {a_1}^2 a'^2 -2ra {a_1}^2 a'' + \left( 2 a^2 \left( 3a_1-1 \right) -3 a_1 a' a r \right) a'_1 }
{2a_1 a \left\{ a_1 a' r-2 a \left( a_1-1 \right) \right\} }{dr}} \, .
\end{align}
Here $c_1$ and $c_2$ are constants of the integration.

Let us assume the $r$-dependencies of $\rho$ and $a_1$, $\rho=\rho(r)$ and $a_1=a_1(r)$.
Then by using the EoS $p=p(\rho)$, we find the $r$-dependence of $p$, $p=p(r)=p\left( \rho\left(r\right) \right)$.
Furthermore by using (\ref{FRN3}), we find the $r$-dependence of $a$, $a=a(r)$.
However, Eq.~(\ref{FRN3}) is not valid outside the compact star because $\rho$ and $p$, of course, vanish there.
Then outside the compact star, we may properly assume the profile of $a(r)$ so that $a(r)$ and $a'(r)$ are continuous
at the surface, that is, the boundary of the compact star, and coincide with $a(r)$ and $a'(r)$ obtained from (\ref{FRN3}).
Therefore by using (\ref{f3}), we find the $r$-dependence of $f$, $f=f(r)$ and by using Eqs.~(\ref{V2}) and (\ref{xi2}),
we find the $r$ dependencies of $V$ and $\xi$, $V=V(r)$ and $\xi=\xi(r)$.
By solving $\xi=\xi(r)$ with respect to $r$, $r=r(\xi)$, we find $f$ and $V$ as functions of $\xi$,
$f(\xi)=f\left( r \left( \xi \right) \right)$, $V(\xi)=V\left( r \left( \xi \right) \right)$ which realize the model which has a solution
given by $\rho=\rho(r)$ and $a_1=a_1(r)$.

We should note, however, the expression of $\xi$ in (\ref{xi2}) gives a constraint,
\begin{align}
\label{cons3}
\frac{2}{a_1} \left(\rho + p \right)
 + {\frac{8\,a a_1 \left(a_1 -1 \right)f'' - \left[ 4 \left( 1-3\,a_1 \right) f' +r \right] \left(a a'_1 -a_1 a' \right) }{a_1 r^2a }} \geq 0 \, ,
\end{align}
so that the ghost could be avoided.
If Eq.~(\ref{cons3}) is not satisfied, the scalar field $\xi$ becomes pure imaginary.
We may define a new real scalar field $\zeta$ by $\xi=i\zeta$ $\left(i^2=-1\right)$ but because the coefficient
in front of the kinetic term of $\zeta$ becomes negative, $\zeta$ is a ghost, that is, a non-canonical scalar field.
The existence of the ghost generates the negative norm states in the quantum theory and therefore
the theory becomes inconsistent.

When we consider compact stars like neutron stars, we often consider the following equation of state,
\begin{enumerate}
\item Energy-polytrope
\begin{align}
\label{polytrope}
p = K \rho^{1 + \frac{1}{n}}\,,
\end{align}
with constants $K$ and $n$.
It is known that for the neutron stars, $n$ could take the value $0.5\leq n \leq 1$.
\item Mass-polytrope
\begin{align}
\label{MassPolytropicEOS}
\rho = \rho_{m} + N p \, ,\qquad \qquad p = K_m \rho_m^{1+\frac{1}{n_{m}}} \, ,
\end{align}
where $\rho_{m}$ is rest mass energy density and $K_{m}$, $N$ are constants,
\end{enumerate}

Now let us study the case of the energy-polytrope (\ref{polytrope}), in detail, in which we can rewrite the EoS as follows,
\begin{align}
\label{polytrope2}
\rho = \tilde K p^{({1 + \frac{1}{\tilde n}})}\, , \quad
\tilde K \equiv K^{-\frac{1}{1+\frac{1}{n}}} \, , \quad
\tilde n \equiv \frac{1}{\frac{1}{1+\frac{1}{n}} - 1}
= - 1 - n \, .
\end{align}
For the energy-polytrope, Eq.~(\ref{FRN3}) takes the following form,
\begin{align}
\label{FRN3p1B}
\frac{1}{2} \ln a = - \int^{p(r)}\frac{dp}{\tilde K p^{1 + \frac{1}{\tilde n}} + p}
= \frac{c}{2} + \tilde n \ln \left(1+{\tilde K}^{-1}p^{-\frac{1}{\tilde n}} \right)
= \frac{c}{2} - \left(1+n\right) \ln \left(1+ K \rho^\frac{1}{n} \right) \, .
\end{align}
Here $c$ is a constant of the integration.
Similarly, in case of mass-polytrope (\ref{MassPolytropicEOS}), we obtain
\begin{align}
\label{masspolytope}
\frac{1}{2} \ln a = \frac{\tilde c}{2} + \ln \left( 1 - K_m \rho^\frac{1}{n_m}\right) \, .
\end{align}
Here $\tilde c$ is a constant of the integration, again.

Under one of the above equations of state, we may assume the following profile of $\rho=\rho(r)$ and $a_1=a_1(r)$,
just for an example,
\begin{align}
\label{anz1}
\rho=\left\{ \begin{array}{cc}
\rho_c \left( 1 - \frac{r^2}{{R_s}^2} \right) & \ \mbox{when}\ r<R_s \\
0 & \ \ \mbox{when}\ r>R_s
\end{array} \right. \, , \quad
a_1= 1 - \frac{2Mr^2}{r^3 + {r_0}^3} \, .
\end{align}
Here $r_0$ is a constant, $\rho_c$ is a constant expressing the energy density at the center of the compact star,
and $R_s$ is also a constant corresponding to the radius of the surface of the compact star, and $M$ is a constant
corresponding to the mass of the compact star,
\begin{align}
\label{MRs}
M =4\pi \rho_c \int_0^r \psi^2 \rho(\psi) d\psi= 4\pi \rho_c \int_0^r d\psi \psi^2 \left( 1 - \frac{\psi^2}{{R_s}^2} \right)
= \frac{4\pi \rho_c r^3}{15}\left( 5 - \frac{3r^2}{{R_s}^2} \right)\,.
\end{align}
When $r\to \infty$, $a_1$ behaves as $a_1(r) \sim 1 - \frac{2M}r$ and therefore $M$ can be regarded as the mass of the compact star.
Eq.~(\ref{MRs}) gives $M$-$r$ relation, that is, the relation between the mass and the radius of the compact star when $r=R_s$.
We also note that we need to choose $r_0$ large enough so that $a_1$ is positive.
In order that $a_1$ in (\ref{anz1}) should be positive, we require
\begin{align}
\label{cond2}
\frac{2^\frac{5}{3}M}{3r_0}<1\, .
\end{align}
We should also note that when $r\to 0$, $a_1$ behaves as $a_1(r) \sim 1 - \frac{2Mr^2}{r_0^3}$\footnote{It is well known that
the junction conditions for the matching of two spacetime manifolds have further restrictions in the EGBS gravity \cite{Davis:2002gn}.
With regards to a static configuration, this { is not a real} problem since the interior will match to vacuum,
and so the pressure will still vanish at the surface $r=R_s$ as we will show below.}.
Therefore $a_1'(r)$ vanishes at the center $r=0$, $a_1'(r=0)=0$, and therefore, there is no conical singularity.

As an example, we use the energy-polytope as the equation of state by choosing $n=1$ just for simplicity.
Then Eq.~(\ref{FRN3p1B}) gives,
\begin{align}
\label{FRN3p1BC1}
a = \frac{\e^c}{ \left( 1 + K \rho_c \left( 1 - \frac{r^2}{{R_s}^2} \right) \right)^4} \, ,
\end{align}
which gives
\begin{align}
\label{FRN3p1BC2}
a' = \frac{8 \e^c K \rho_c r}{{R_s}^2\left( 1 + K \rho_c \left( 1 - \frac{r^2}{{R_s}^2} \right) \right)^5} \, ,
\end{align}
Outside the star, we assume $a(r)=a_1(r)$ in (\ref{anz1}) and therefore
\begin{align}
\label{arprimeout}
a' = \frac{2Mr\left( r^3 - {r_0}^3\right)}{\left( r^3 + {r_0}^3\right)^2} \, .
\end{align}
Because $a(r)$ and $a'(r)$ should be continuous at the surface $r=R_s$. we obtain
\begin{align}
\label{matching}
\e^c = 1 - \frac{2M{R_s}^2}{{R_s}^3 + {r_0}^3} \, , \quad
\frac{8 \e^c K \rho_c}{R_s} = \frac{2M{R_s}\left( {R_s}^3 - {r_0}^3\right)}{\left( {R_s}^3 + {r_0}^3\right)^2} \, .
\end{align}
By deleting $\e^c$ in the two equations in (\ref{matching}), we obtain,
\begin{align}
\label{r0eq}
0=& \left({r_0}^3\right)^2 + \left( 2 {R_s}^3 - 2M{R_s}^2 + \frac{M{R_s}^2}{2K \rho_c} \right){r_0}^3 + {R_s}^6 - 2M{R_s}^5 - \frac{M{R_s}^5}{4K \rho_c} \nonumber \\
=& \left({r_0}^3\right)^2 + \left( 2 {R_s}^3 - \frac{16\pi \rho_c R_s^5}{15} + \frac{2\pi R_s^5}{15K} \right){r_0}^3
+ {R_s}^6 - \frac{16\pi \rho_c R_s^8}{15} - \frac{2\pi R_s^8}{15K} \, ,
\end{align}
where we have used Eq.~(\ref{MRs}) when $r=R_s$.
Because $r_0$ should be positive, we find
\begin{align}
\label{cond}
& {R_s}^6 - \frac{16\pi \rho_c R_s^8}{15} - \frac{4\pi R_s^8}{15K} <0 \nonumber \\
& \mbox{or} \nonumber \\
& 2 {R_s}^3 - \frac{16\pi \rho_c R_s^5}{15} + \frac{4\pi R_s^5}{15K} <0\quad \mbox{and} \quad
{R_s}^6 - \frac{16\pi \rho_c R_s^8}{15} - \frac{4\pi R_s^8}{15K} > 0 \, .
\end{align}
Then by using (\ref{f3}), we find the $r$-dependence of $f$, $f=f(r)$ and by using Eqs.~(\ref{V2}) and (\ref{xi2}),
the $r$ dependencies of $V$ and $\xi$, $V=V(r)$ and $\xi=\xi(r)$, are determined.
If we can solve $\xi=\xi(r)$ with respect to $r$, $r=r(\xi)$, we find $f$ and $V$ as functions of $\xi$,
$f(\xi)=f\left( r \left( \xi \right) \right)$, $V(\xi)=V\left( r \left( \xi \right) \right)$.

Just for further simplicity, we may choose
\begin{align}
\label{smpl}
2 r_0 = R_s = 4 M= \frac{32\pi \rho_c R_s^3}{15}\, , \quad K \rho_c = \frac{7}{258} \, , \quad \e^c= \frac{5}{9}\, ,
\end{align}
which satisfy Eqs.~(\ref{cond2}), (\ref{matching}), and (\ref{r0eq}).
For numerical calculation, we may further choose $R_s=1$.

Inside the compact star, by using Eqs.~(\ref{anz1}) and (\ref{FRN3p1BC1}), we find the GB term $\mathcal{G}$ behaves as,
\begin{align}
\label{GB}
\mathcal{G}(r)=&\, -\frac{K\rho_cM }{64 \left( {R_s}^2+K\rho_c {R_s}^2-K\rho_cr^2 \right)^2
\left( r^3 + {r_0}^3 \right)^3}
\left\{ 9r^5{r_0}^3K\rho_c +3{r_0}^6K\rho_cr^2-4{r_0}^3Mr^4K\rho_c +r^5MK\rho_c{R_s}^2 \right. \nonumber\\
& +3r^3{r_0}^3K \rho_c {R_s}^2 +3{r_0}^6K\rho_c{R_s}^2 -8{r_0}^3Mr^2K\rho_c{R_s}^2 +r^5 M {R_s}^2
+6r^8K\rho_c -13r^7MK\rho_c \nonumber\\
& \left. +3r^3{r_0}^3{R_s}^2 +3{r_0}^6{R_s}^2-8{r_0}^3 Mr^2{R_s}^2 \right\} \nonumber \\
\approx&\, - \frac{64{R_s}^2{r_0}^3 MK\rho_c}{{r_0}^6{R_s}^4 \left( K\rho_c+1 \right) }
+ \frac{64 \left( 8M{R_s}^2 +8MK\rho_c{R_s}^2 -9{r_0}^3 K\rho_c \right) MK\rho_cr^2}
{{r_0}^6{R_s}^4 \left( K\rho_c+1 \right)^2} +\frac{384MK\rho_c r^3}{{R_s}^2{r_0}^6 \left( 1+K\rho_c \right)}\,.
\end{align}
Eq.~(\ref{GB}) shows that the GB term does not vanish and it depends on the Mass of the star.
Now we calculate the form of $f(r)$ by using the data given in Eqs.~(\ref{FRN3p1BC1}), (\ref{matching}), and (\ref{anz1}).
The explicit form of $f(r)$ is displayed in Appendix A.
The form of $\xi(r)$, by using the data given in Eqs.~(\ref{anz1}), (\ref{FRN3p1BC1}), and (\ref{matching}), is also displayed in Appendix A.
Finally, we calculate the explicit form of $V(r)$ by using the data given in Eqs.~(\ref{FRN3p1BC1}), (\ref{matching}), and (\ref{anz1}) and list the results in Appendix A. 

{ To complete our study, we solve Eq.~(\ref{A4}) asymptotically and obtain, }
\begin{align}
\label{rxi}
\xi(r\rightarrow 0) \approx& C_{11}+C_{12}\sqrt{r} \ \Rightarrow\ r \approx C_{13}+C_{14}\xi+C_{15}\xi^2\, .
\end{align}
The above equation is valid provided that the constant $C_{12}<0$.
Now using Eq.~(\ref{rxi}) in (\ref{A2}), we obtain $f(\xi)$ as
\begin{align}
f(\xi)\approx C_{16}+C_{17}\xi +C_{18}\xi^2\,.
\end{align}
Also using Eq.~(\ref{rxi}) in (\ref{A6}), we obtain $V(\xi)$ as:
\begin{align}
V(\xi)\approx C_{19}+C_{20}\xi+C_{21}\xi^2\,.
\end{align}
A final { remark} that we should stress is the fact that the use of Eqs.~(\ref{anz1}), (\ref{FRN3p1BC1}), (\ref{A2}), and the constraints (\ref{smpl})
with $R_s=1$, one can show easily that the inequality (\ref{cons3}) is hold.

We have four differential equations for seven unknown functions, as shown in Eqs.~\eqref{Eq2tt}, \eqref{Eq2rr}, \eqref{Eq2pp}, and \eqref{Eqphi2},
that is $\rho$, $p$, $V$, $\xi$, $f$, $a$, and $a_1$.
As a result, we need to require three additional conditions to close such a system.
One of these extra conditions is the continuity equation given by Eq.~(\ref{FRN2}).
The second condition is the polytropic equation of state given by Eq.~(\ref{polytrope}).
The third one is the profile of the energy density of matter given by Eq.~\eqref{anz1}.
When these additional conditions are combined with Eqs.~\eqref{Eq2tt}, \eqref{Eq2rr}, \eqref{Eq2pp}, and \eqref{Eqphi2}, the system is in a closed form,
allowing all seven unknown functions to be explicitly fixed. 

\section{Ingredient requirements for a real physical stellar }\label{S3}

For a physically reliable isotropic stellar model, the solution has to satisfy the below-listing conditions inside the stellar configurations,
\begin{itemize}
\item The metric potentials $a(r )$ and $a_1(r )$, and the energy-momentum components $\rho$ and $p$ should be well defined at the
center of the star and have a regular behavior and have no singularity in the interior of the star.
\item The density $\rho$ must be positive in the stellar interior i.e., $\rho\geq 0$.
Moreover, its value at the center of the star must be finite, positive, and decreasing to the boundary of the stellar i.e., $\frac{d\rho}{dr}\leq 0$.
\item The pressure $p$ should have the positive value inside the fluid configuration i.e., $p\geq0$.
In addition, the derivative of the pressure should yield a negative value inside the stellar, i.e., $\frac{dp}{dr}< 0$.
At the surface of the stellar, $r= R_s$, the pressure $p$ should vanish.
\item For an isotropic fluid sphere, the inquiries of the energy conditions are given by the following inequalities in every point:
\begin{enumerate}
\item Null energy condition (NEC): $\rho> 0$.
\item Weak energy condition (WEC): $p+\rho > 0$.
\item Strong energy condition (SEC): $\rho+3p > 0$.
\end{enumerate}
\item The causality condition which should be satisfied to { have} a realistic model, i.e. the speed of sound should be less than $1$
(provided that the speed of light is $c = 1$) in the interior of the star, i.e., { $1\geq\frac{dp}{d\rho}\geq 0$}.
\item To have a stable model, the adiabatic index must be more than $\frac{4}{3}$.
\end{itemize}
It is time to analyze the above conditions to see if we have a real isotropic star or not.

\section{Physical behavior of our model}\label{S4}

To test if our model given by Eqs.~(\ref{MassPolytropicEOS}) and (\ref{FRN3p1B}) agrees with a real stellar construction, we discuss the following issues:

\subsection{Non singular model}

\begin{enumerate}
\item The metric potentials of this model satisfy,
\begin{align}
\label{sing}
a(r\rightarrow 0)=\frac{\e^c}{\left(1+K\rho_c\right)^4}\quad \textrm{and} \quad a_1(r\rightarrow 0)=1\, ,
\end{align}
that yields that the metric potentials have finite values at the center of the star configuration.
Additionally, the derivatives of these metric potentials vanish at the center of the star, i.e., $a'(r\rightarrow0)=a'_1(r\rightarrow0)=0$.
If the derivatives do not vanish even finite, there appear conical singularities at the center.
The above constraints { yield} that the metric is regular at the center as well as { the metric} has a good
behavior in the interior of the stellar.
\item Density (\ref{anz1}) and pressure (\ref{polytrope}), at the center, have the form
\begin{align}
\label{reg1}
\rho(r\rightarrow 0)=\rho_c\, , \quad p(r\rightarrow 0)=K{\rho_c}^2\, .
\end{align}
The above Eq.~(\ref{reg1}) shows clearly that the density and pressure at the center of the star have always positive values
if $\rho_c>0$ and $K>0$ otherwise they become negative.
\item The gradient of density and pressures of our model are given respectively as
\begin{align}
\label{dsol}
\rho'=-\frac{2\rho_c r}{{R_s}^2}\, , \quad p'=-\frac{4K\rho_c r \left(1-\frac{r^2}{{R_2}^2}\right)}{{R_2}^2}\, .
\end{align}
Here $\rho'=\frac{d\rho}{dr}$ and $p'=\frac{dp}{dr}$.
Equation~(\ref{dsol}) shows that the derivatives of density and pressure are negative.
Furthermore, because they vanish at the center of the star, the conical singularities do not appear.
\item The velocity of sound using relativistic units, i.e., $\left(c=G=1\right)$ are derived as \cite{HERRERA1992206},
\begin{align}
\label{dso2}
{v_r}^2=\frac{dp}{d\rho}=\frac{2\rho_c \left({R_s}^2 -r^2 \right)}{{R_s}^2}\, .
\end{align}
\end{enumerate}
Now we are ready to plot all the above conditions to see their behaviors using the numerical { constraints} listed in Eq.~(\ref{smpl}).

\begin{figure}
\centering
\subfigure[~Metric potential $a_1(r)$]{\label{fig:pot2}\includegraphics[scale=.4]{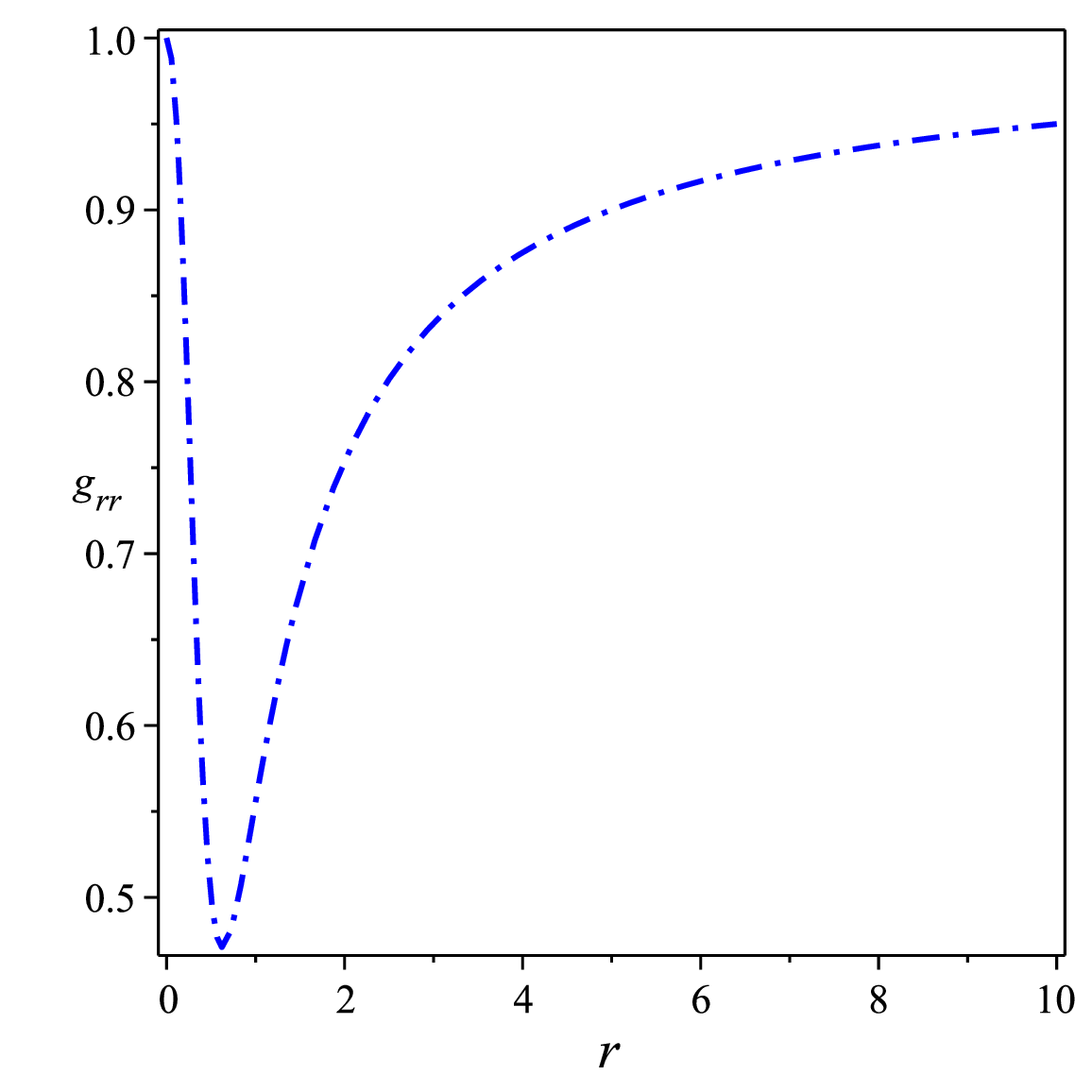}}
\subfigure[~Metric potential $a(r)$]{\label{fig:pot1}\includegraphics[scale=0.4]{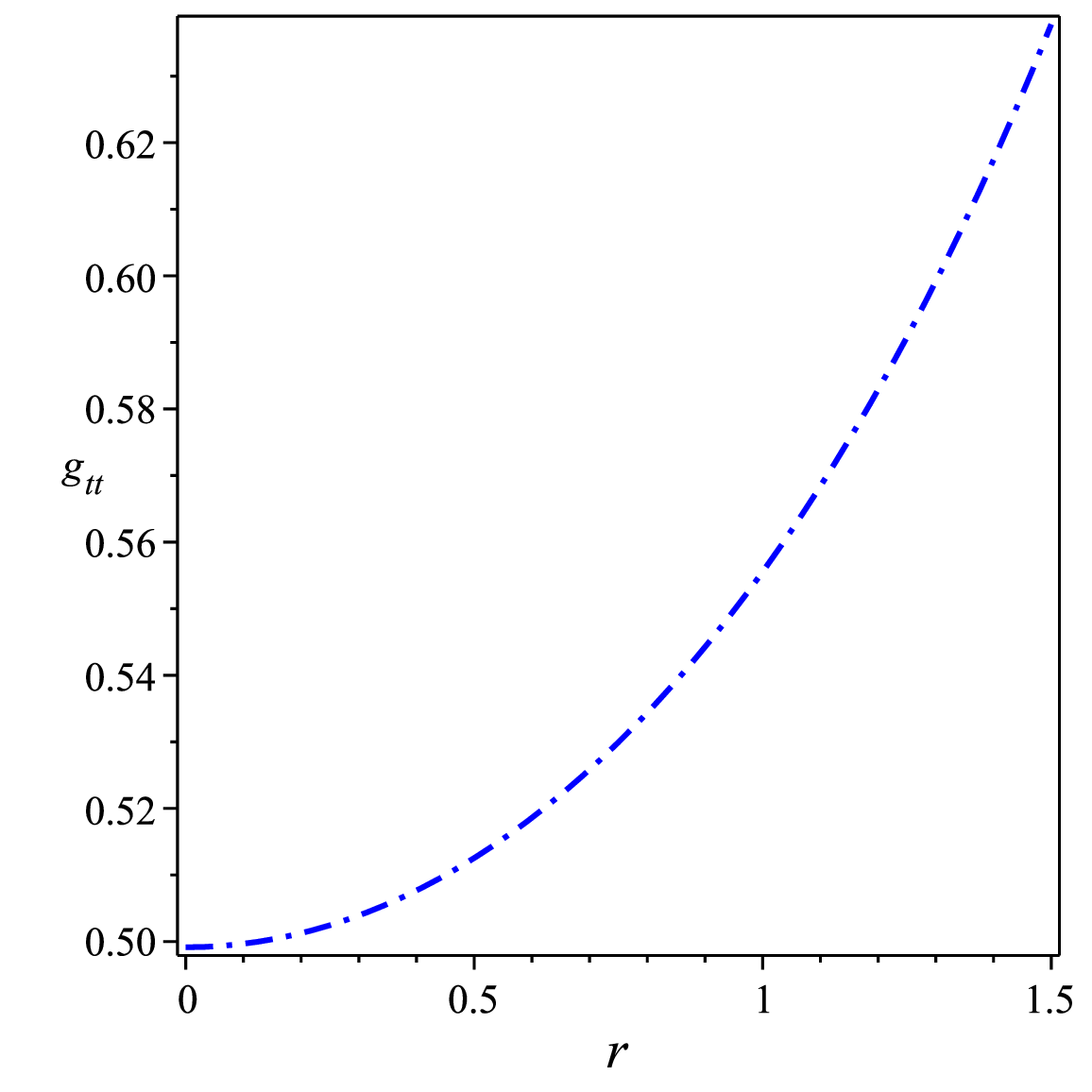}}
\caption[figtopcap]{\small{{Schematic plot of the metric potentials (\ref{anz1}) and (\ref{FRN3p1BC1})
vs. the radial coordinate $r$ using the { constraints}~(\ref{smpl}).}}}
\label{Fig:1}
\end{figure}

In { Figures}~\ref{Fig:1}~\subref{fig:pot2} and \ref{Fig:1}~\subref{fig:pot1}, we present the behavior of metric potentials.
As Figure~\ref{Fig:1} shows, the metric potentials assume the values $a_1(r\rightarrow 0)=1$
and $a(r\rightarrow 0)=0.5$ for $r=0$, which ensure that both of the metric potentials have finite and positive values at the center of the star.

\begin{figure}
\centering
\subfigure[~Density]{\label{fig:density}\includegraphics[scale=0.4]{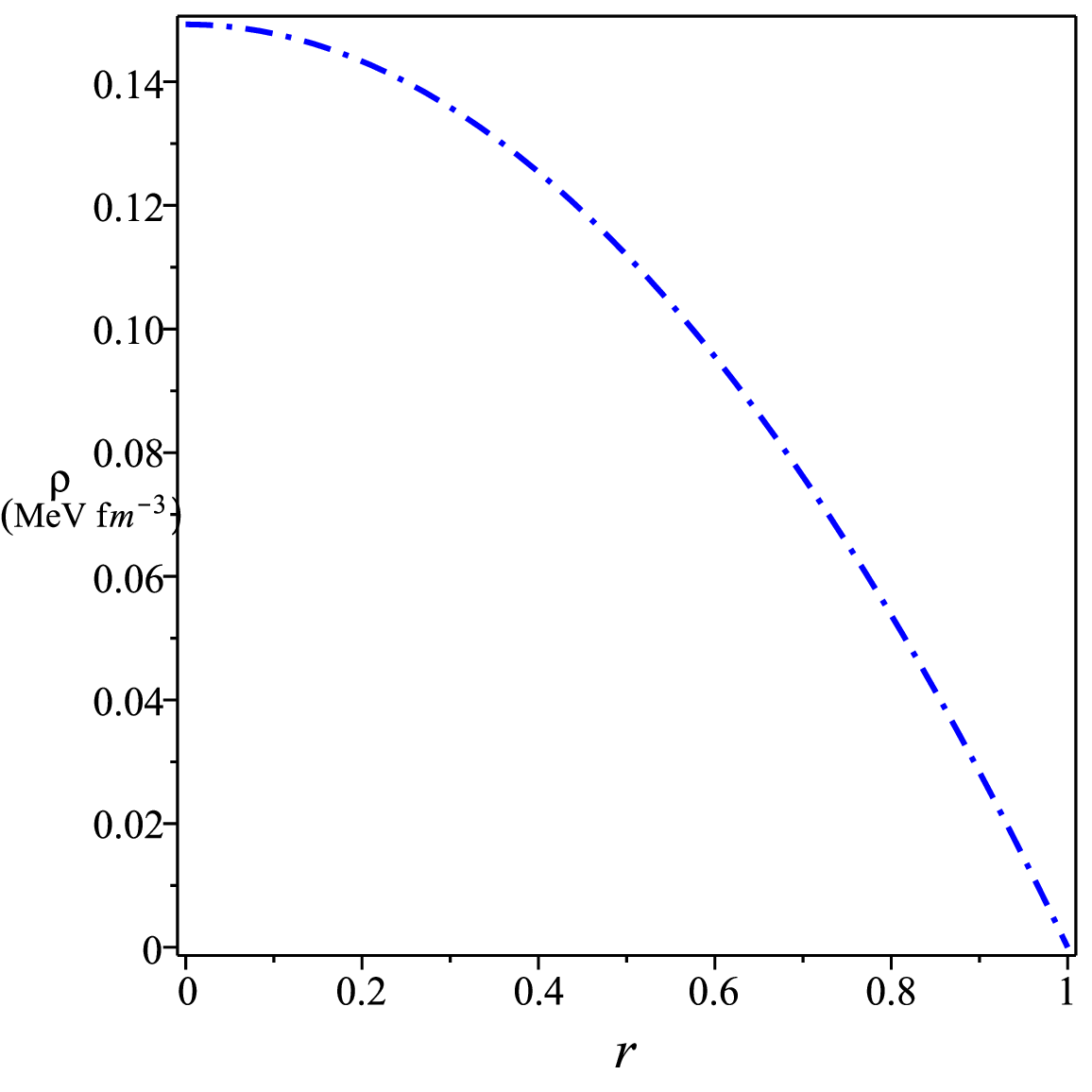}}
\subfigure[~Pressure]{\label{fig:pressure}\includegraphics[scale=.4]{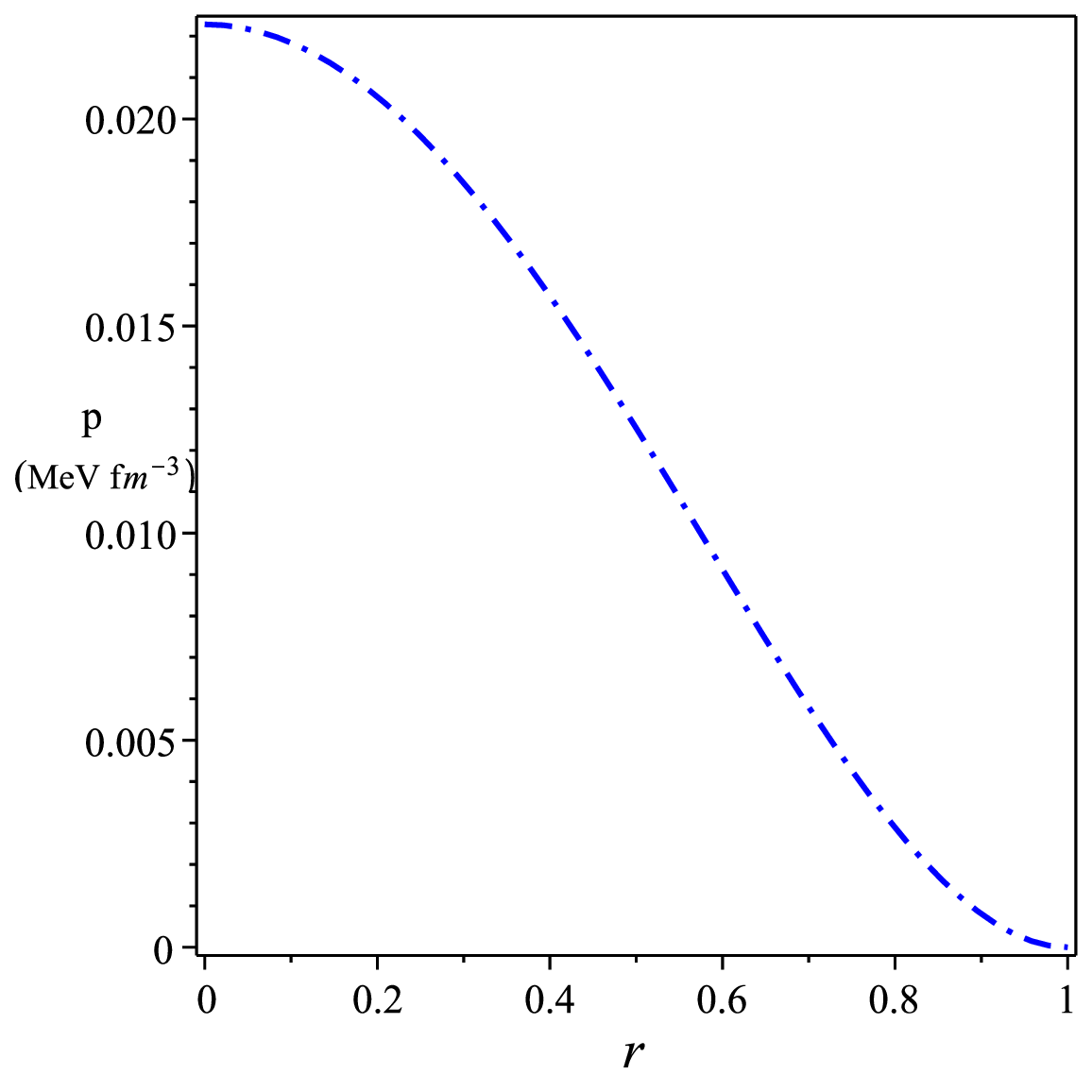}}
\caption[figtopcap]{\small{{{Plot of the { energy density} and pressure of (\ref{polytrope}) and (\ref{anz1}) vs. the radial coordinate $r$
using the constraints (\ref{smpl}).}}}}
\label{Fig:2}
\end{figure}

Now we plot the { energy} density and pressures, listed by Eqs.~(\ref{polytrope}) and (\ref{anz1}) in Figure~\ref{Fig:2}.
As Figure~\ref{Fig:2} shows that { the energy density} and pressure are positive which { is} in agreement for a realistic stellar configuration.
Additionally, as { Figures}~\ref{Fig:2}~\subref{fig:density} and \subref{fig:pressure} indicate, the density and pressure have high values
at the center and decrease toward the boundary, which is relevant for a realistic star.

\begin{figure}
\centering
\subfigure[~Gradient of density]{\label{fig:drho}\includegraphics[scale=0.4]{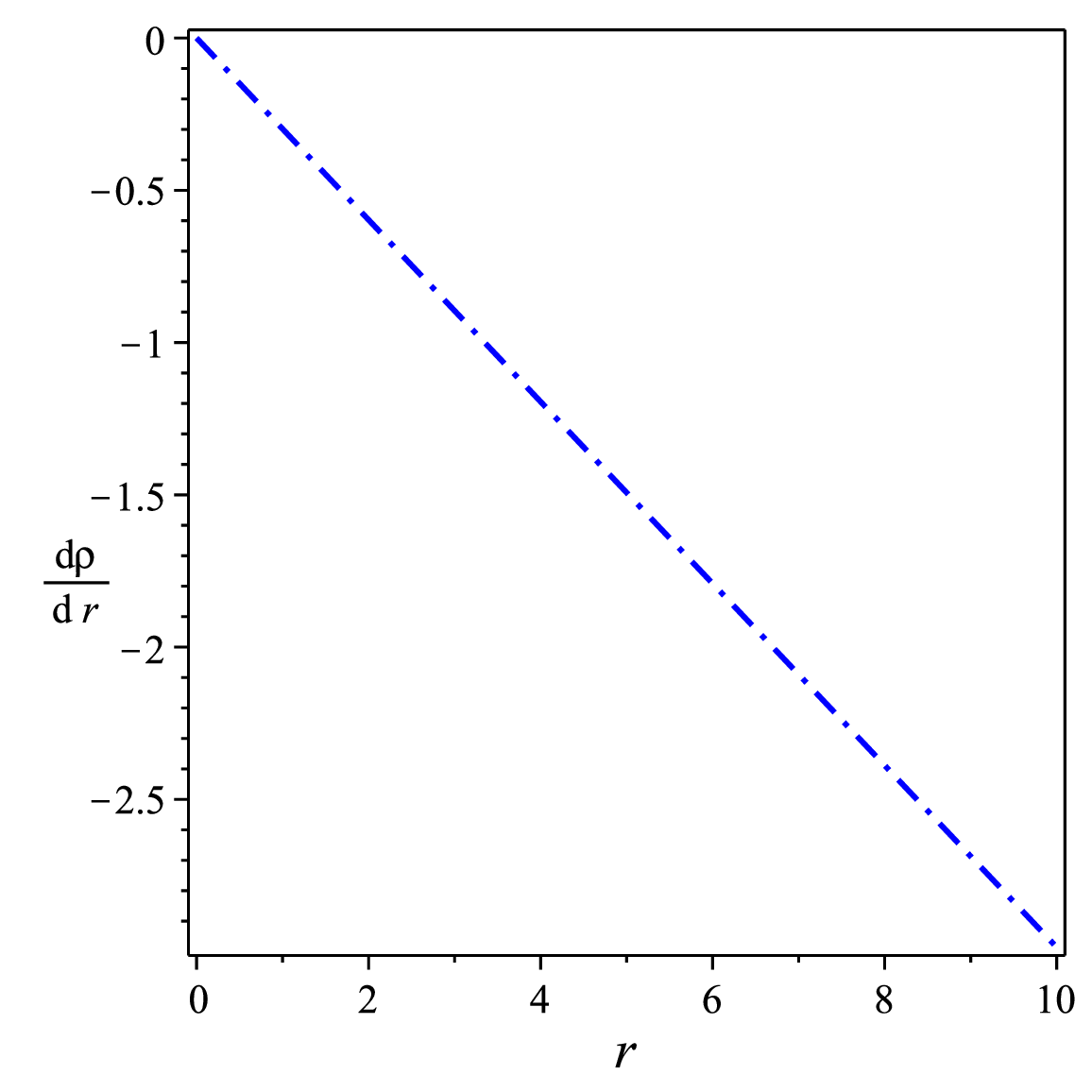}}
\subfigure[~Gradient of pressure]{\label{fig:dp}\includegraphics[scale=.4]{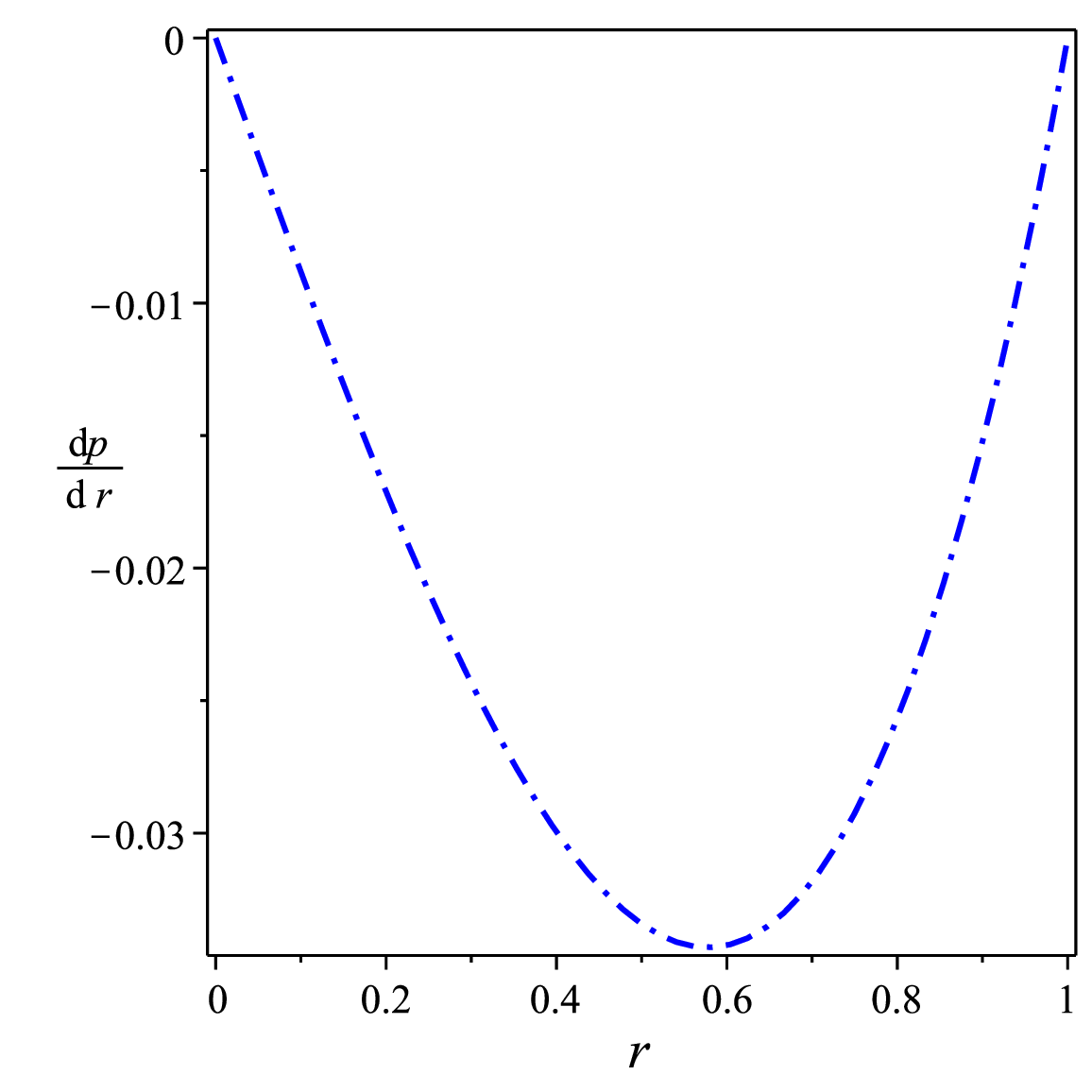}}
\caption[figtopcap]{\small{{{ Plot of the gradients of density and pressure of (\ref{polytrope}) and (\ref{anz1}) vs.
the radial coordinate $r$ using the { constraints}~(\ref{smpl}).}}}}
\label{Fig:3}
\end{figure}

Figure~\ref{Fig:3} shows that the derivatives of density and pressure have negative values,
which ensure the decreasing of density and pressure throughout the stellar configuration.
\begin{figure}
\centering
\subfigure[~Speed of sound]{\label{fig:dpr}\includegraphics[scale=0.3]{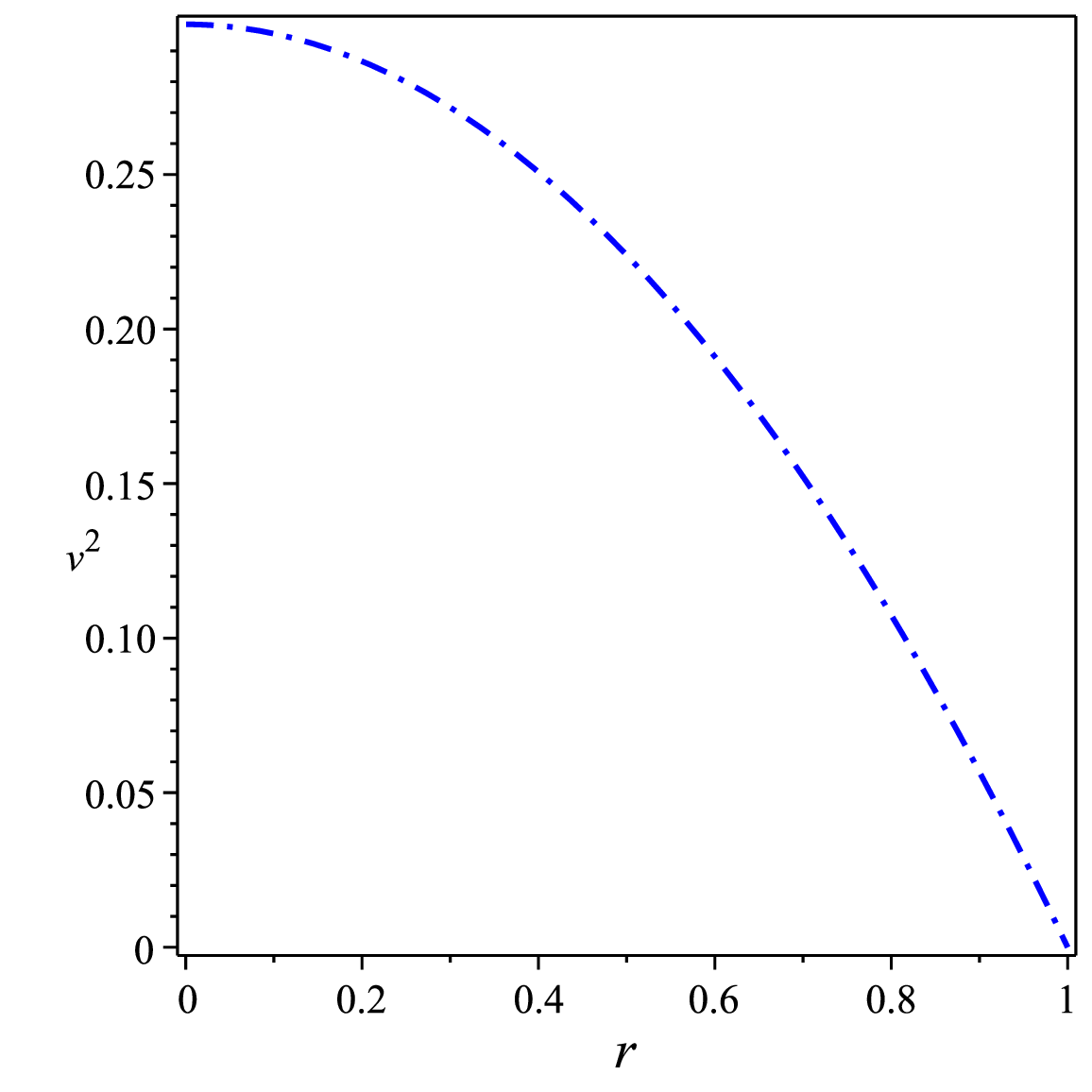}}
\subfigure[~Mass-radius relation]{\label{fig:pressure}\includegraphics[scale=.3]{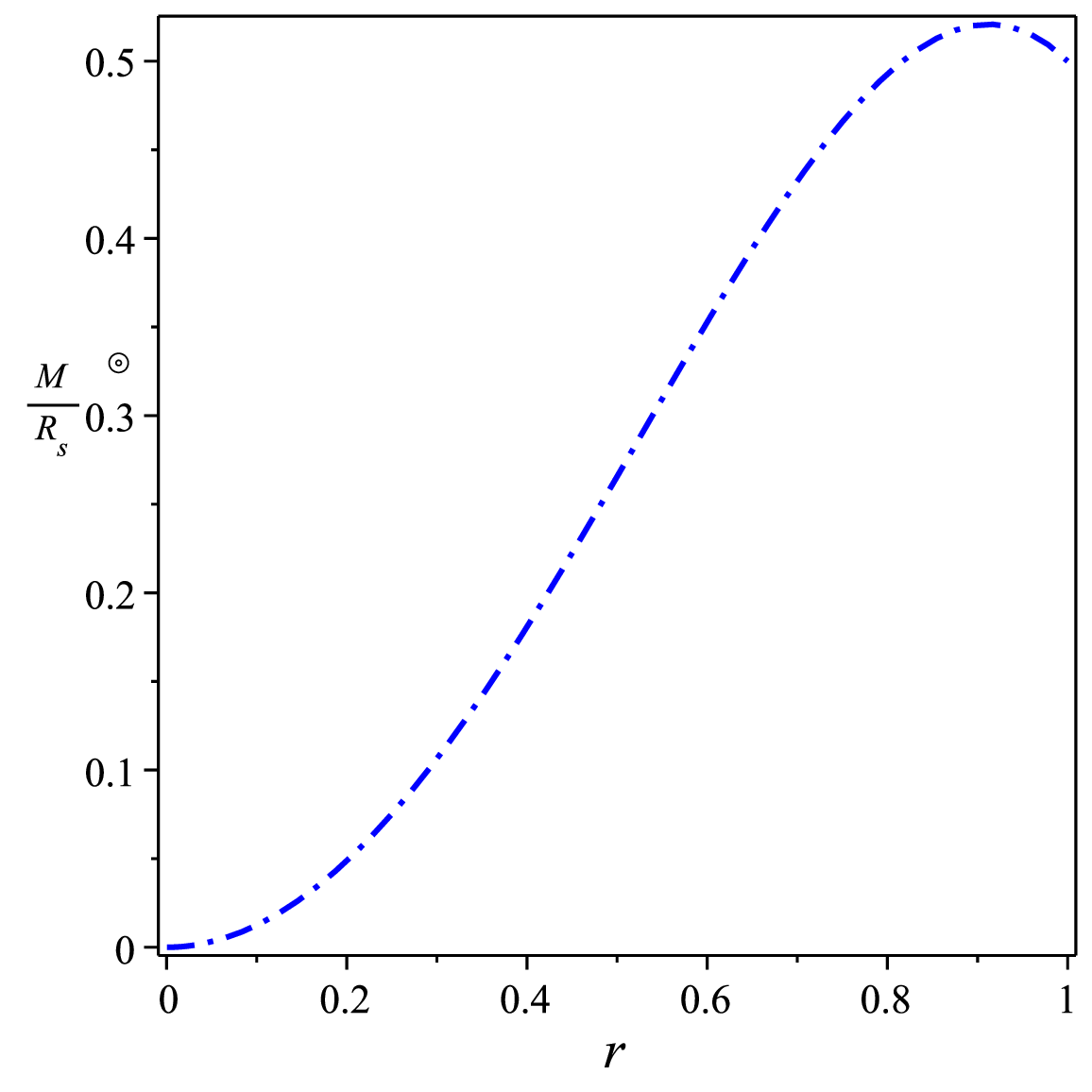}}
\subfigure[~Compactness]{\label{fig:comp}\includegraphics[scale=.3]{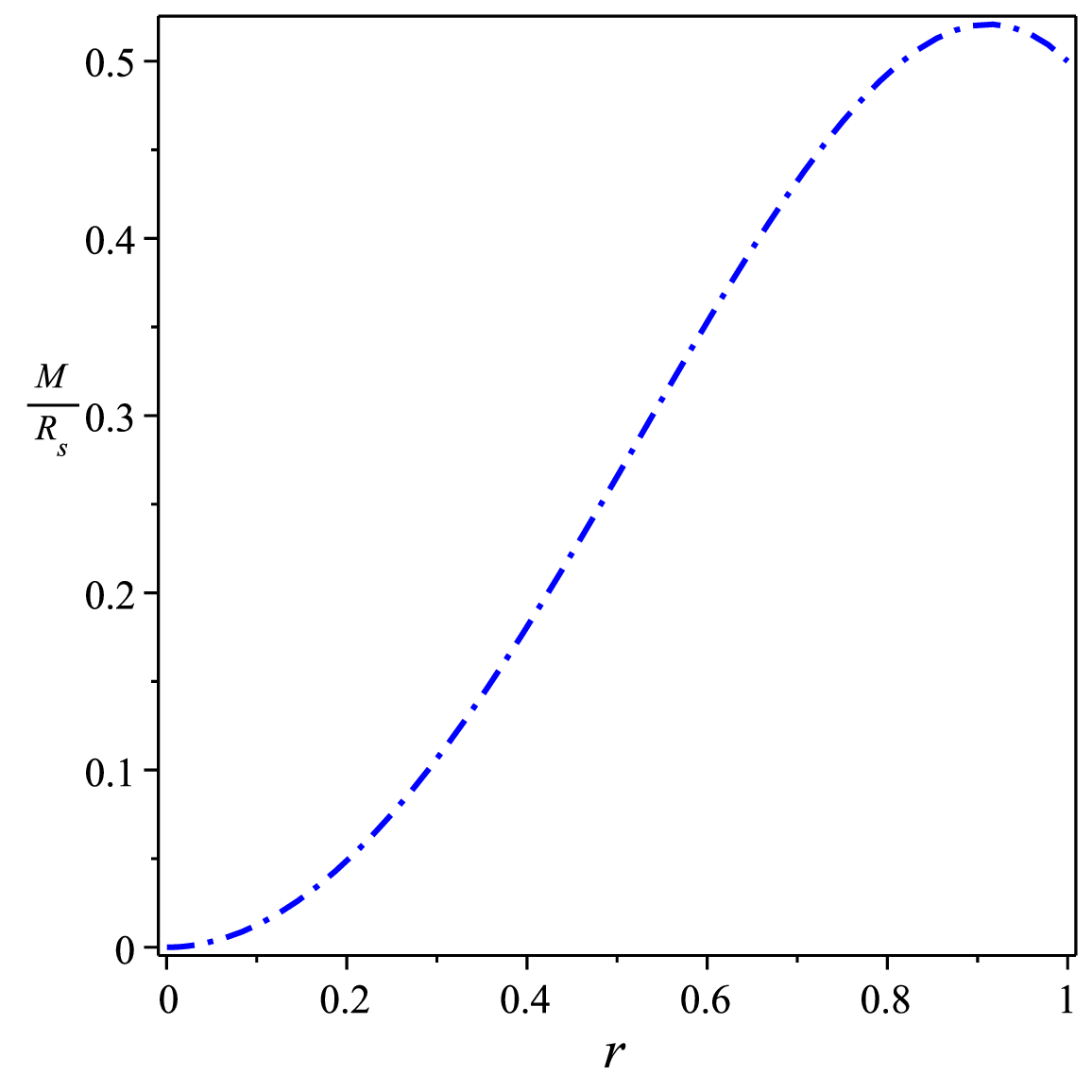}}
\caption[figtopcap]{\small{{{Plot of the speed of sound~\subref{fig:dpr}, mass-radius relation~\subref{fig:pressure}, and compactness
of the stellar~\subref{fig:comp} via the radial coordinate $r$ using the { constraints}~(\ref{smpl}).}}}}
\label{Fig:4}
\end{figure}

In Figure~\ref{Fig:4}, we plot the speed of sound and the mass-radius relation.
As Figure~\ref{Fig:4}~\subref{fig:dpr} { shows, the speed of sound is less than one,}
which confirms the non-violations of causality condition in the interior of the stellar configuration.
Moreover, Figure~\ref{Fig:4}~\subref{fig:comp} shows that the compactness of our model is constrained by $0<C<0.55$,
where $C=\frac M r$ in the stellar configuration. As Figure~\ref{Fig:4}~\subref{fig:dpr} { shows, the causality}
condition is satisfied which is one of the { merits in} this study due to the procedure we follow { although in the frame of GR,}
it is shown this condition is not satisfied \cite{Mak:2013pga}.
We may infer that the procedure used in this study is responsible for the correction in the behavior of the causality condition.
Moreover also as in \cite{Mak:2013pga}, it is shown that the maximum mass lies in the range $0.2 \, M_\odot$.
{ In our model, however, due to the procedure we follow in this study,} the maximum mass is about $0.25\, M_\odot$
as shown in Figure~\ref{Fig:4}~\subref{fig:pressure}, which could be { used} to confront it with the recent data. 
\begin{figure}
\centering
\subfigure[~Null energy conditions]{\label{fig:NEC}\includegraphics[scale=0.3]{JFRMMMAB_den.eps}}
\subfigure[~Weak energy conditions]{\label{fig:WEC}\includegraphics[scale=0.3]{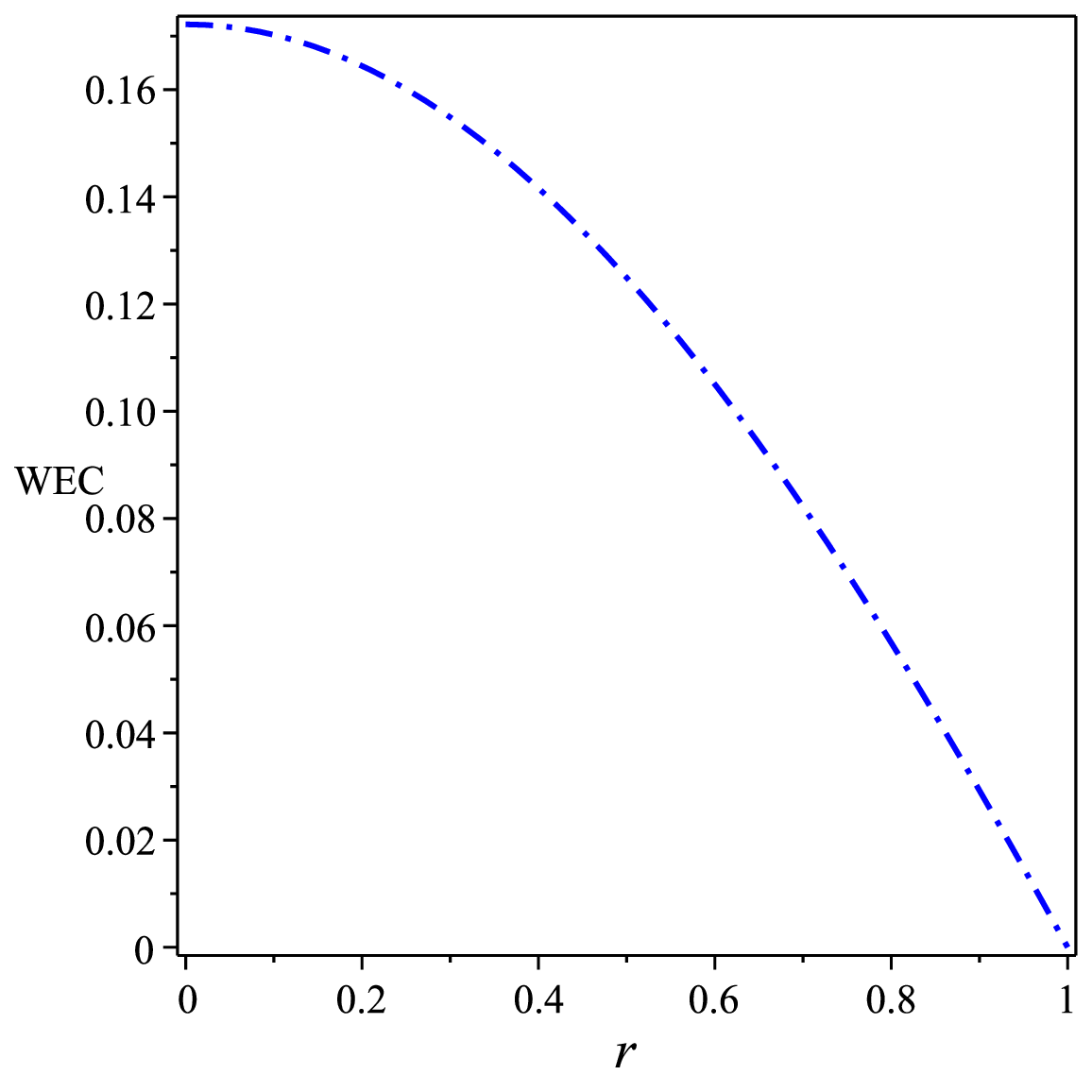}}
\subfigure[~Strong energy condition]{\label{fig:SEC}\includegraphics[scale=.3]{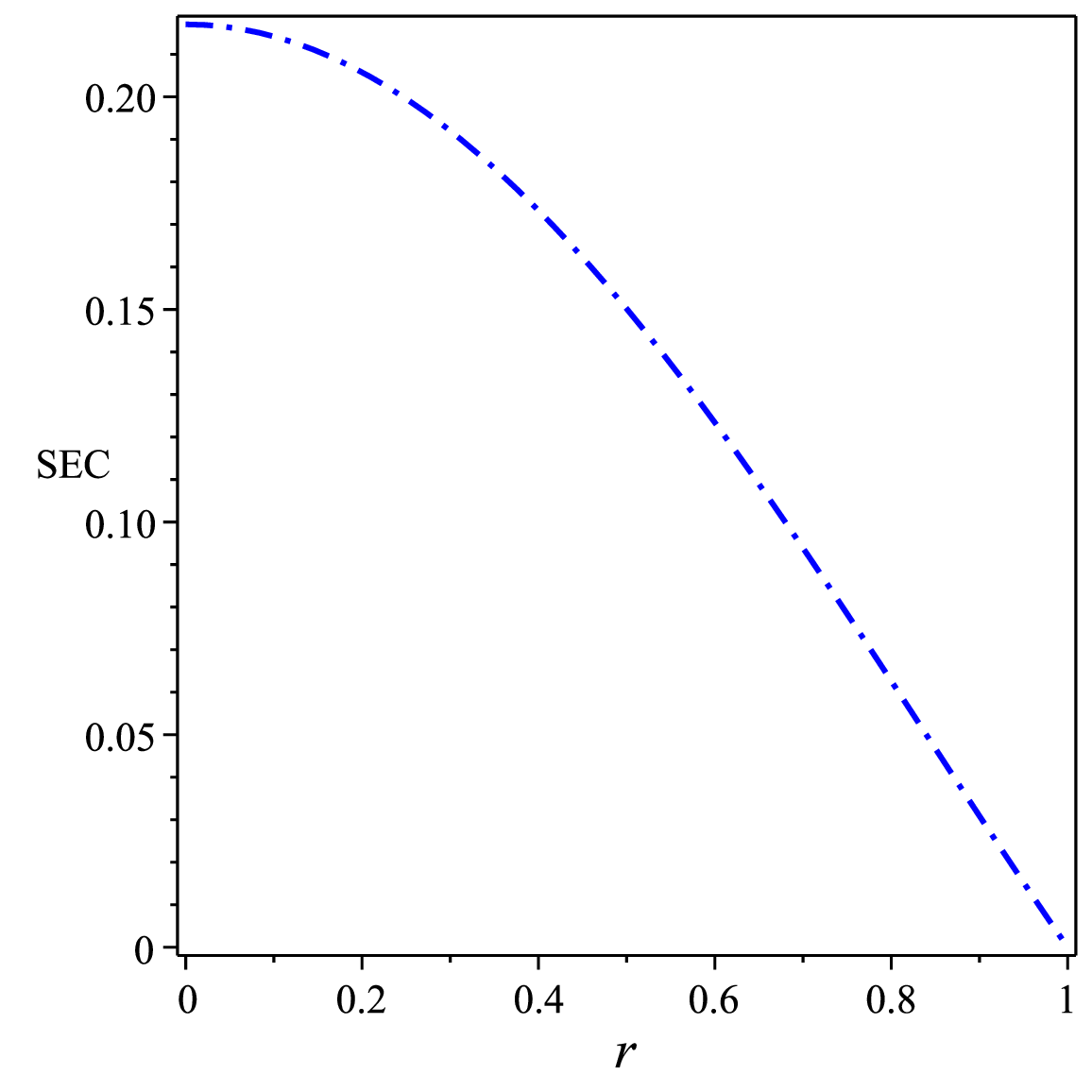}}
\caption[figtopcap]{\small{{{Plot of the null, week, and strong energy conditions of (\ref{polytrope}) and (\ref{anz1}) vs.
the radial coordinate $r$ using the { constraints}~(\ref{smpl}).}}}}
\label{Fig:5}
\end{figure}

Figure~\ref{Fig:5} shows the behavior of the energy conditions.
Particularly, { Figures}~\ref{Fig:5}~\subref{fig:NEC}, \subref{fig:WEC},
and \subref{fig:SEC} indicate the positive values of the NEC, WEC, and SEC energy conditions,
which ensure that all the conditions are verified through the stellar configuration as it should { be}
for a physical stellar model.
\begin{figure}
\centering
\subfigure{\label{fig:EoS}\includegraphics[scale=0.4]{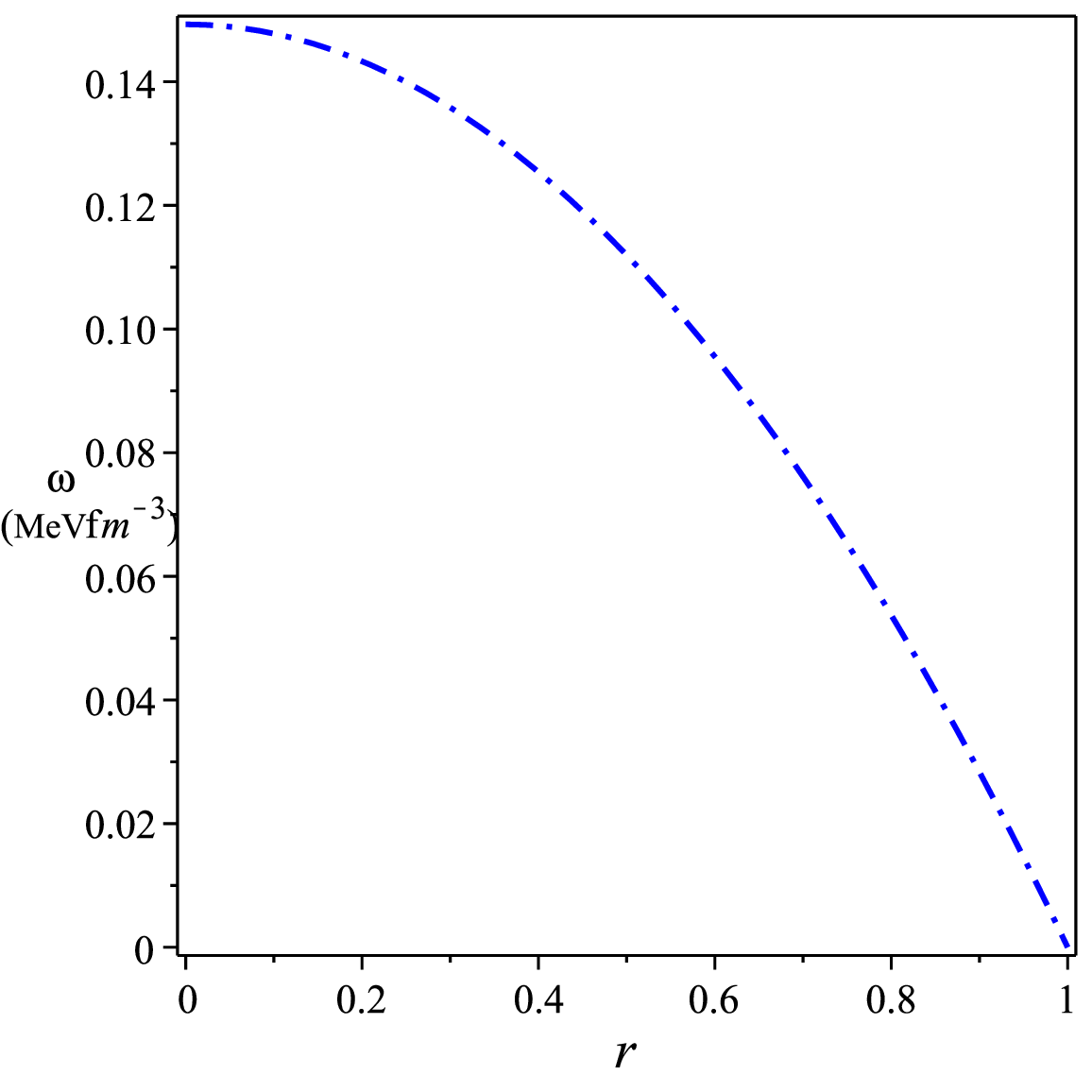}}
\subfigure{\label{fig:Z}\includegraphics[scale=.4]{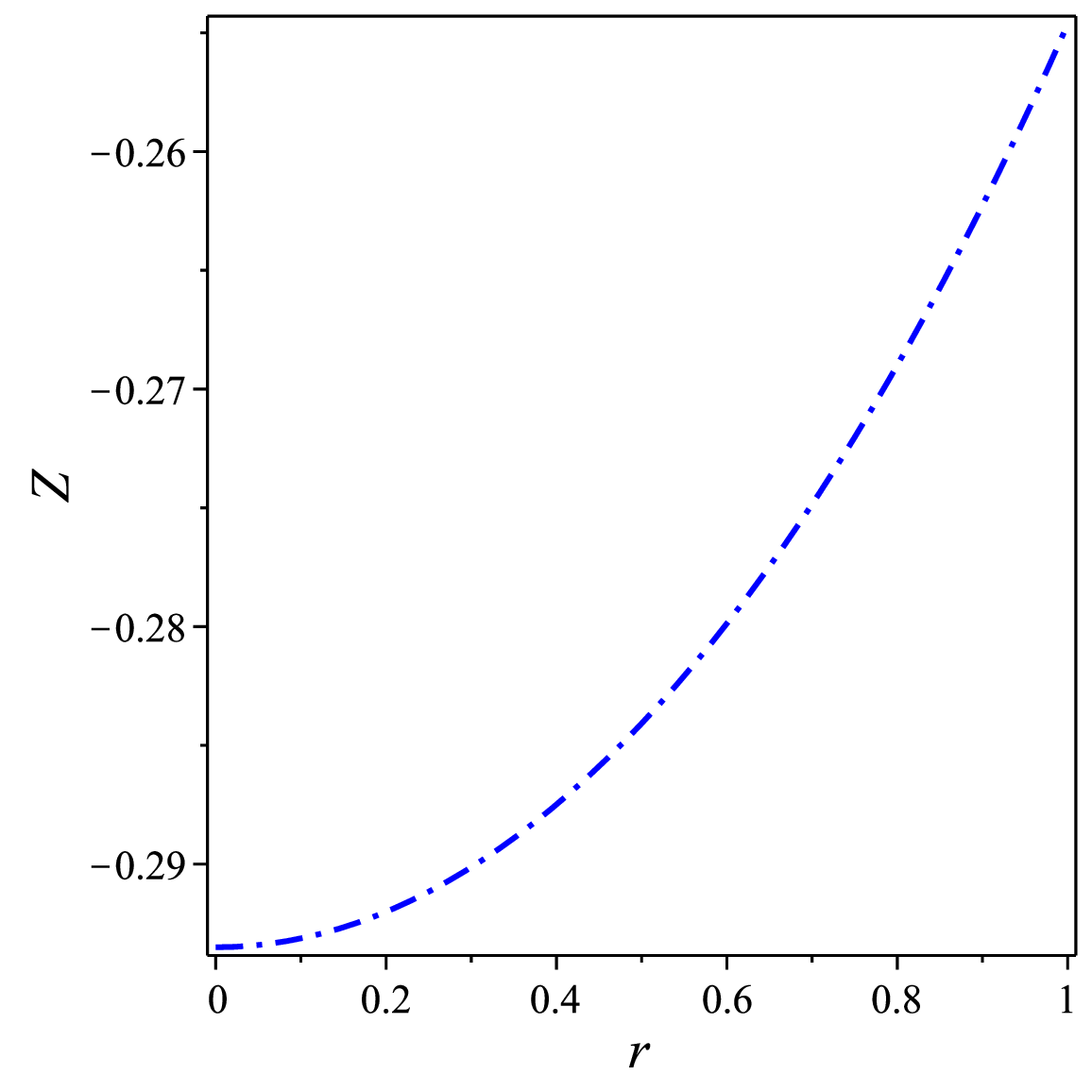}}
\caption[figtopcap]{\small{{Plot of the EoS vs. the radial coordinate $r$~\subref{fig:EoS} and the red shift~\subref{fig:Z} using the { constraints}~(\ref{smpl}).}}}
\label{Fig:6}
\end{figure}

In Figure~\ref{Fig:6}, we plot the EoS.
As { Figure~\ref{Fig:6}~\subref{fig:EoS} shows,} the EoS is not linear.
It was shown in \cite{Das:2019dkn} that the EoS of neutral compact stars is almost a linear one in contrast to the EoS presented
in this study, which shows a non-linear form { due} to the form of the pressure given by Eq.~(\ref{polytrope}).

\section{Stability of the model}\label{stability}

Now we are ready to test the stability issue on our model using the adiabatic index.
The stable equilibrium of a spherically symmetric space-time can be investigated through the adiabatic index which is an ingredient tool
to { test} the stability criterion.
The adiabatic perturbation, i.e., the adiabatic index $\Gamma$, is defined as \cite{1964ApJ...140..417C,1989A&A...221....4M,10.1093/mnras/265.3.533},
\begin{align}
\label{ai}
\Gamma=\left(\frac{\rho+p(r)}{p(r)}\right)\left(\frac{dp(r)}{d\rho(r)}\right)\,.
\end{align}
A Newtonian isotropic sphere has a stable equilibrium if the adiabatic index $\Gamma>\frac{4}{3}$ \cite{1975A&A....38...51H}.
If $\Gamma=\frac{4}{3}$, the isotropic sphere is in neutral equilibrium.

\begin{figure}
\centering
\includegraphics[scale=0.3]{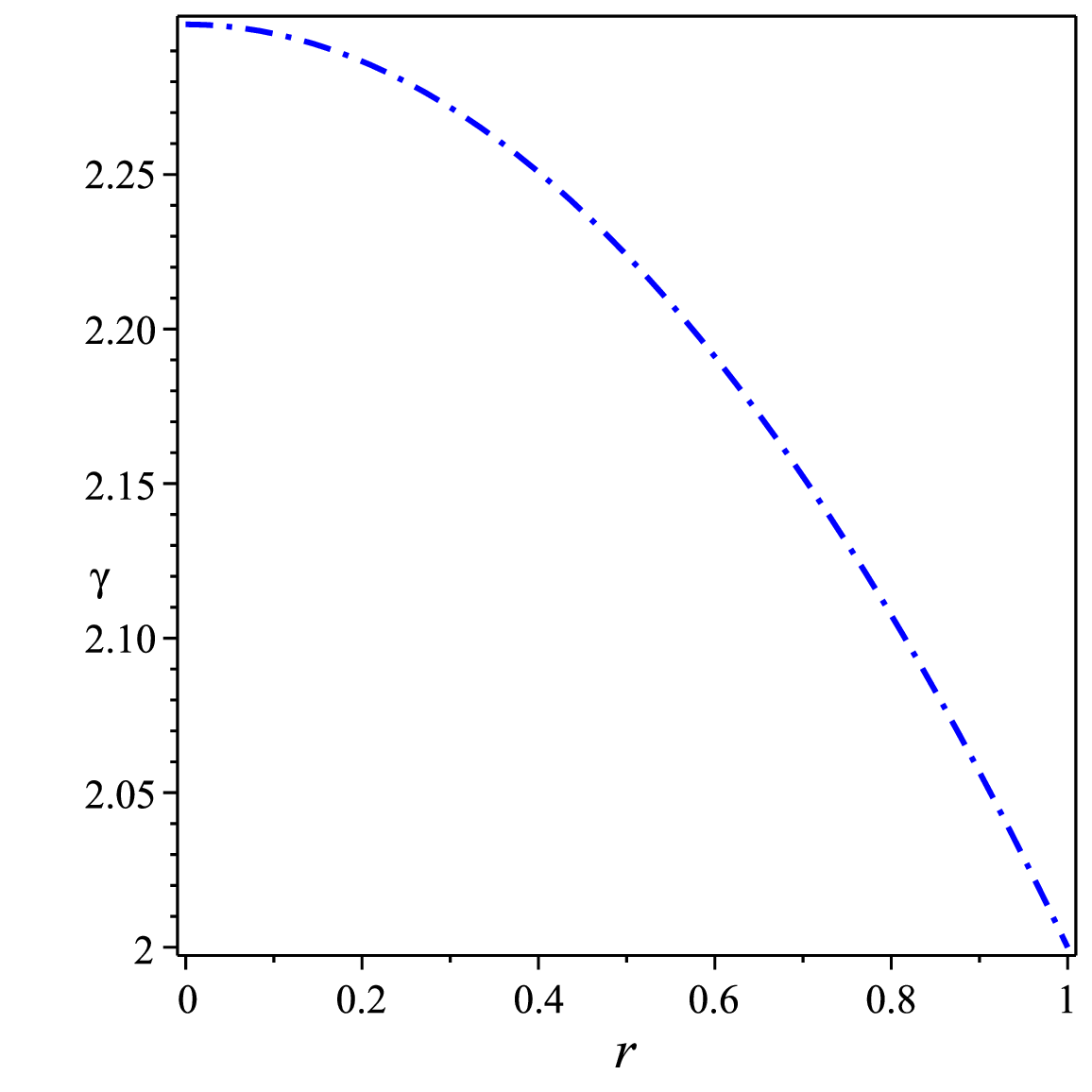}
\caption[figtopcap]{\small{
{Plot of the adiabatic index using the { constraints}~(\ref{smpl}).}}}
\label{Fig:7}
\end{figure}

Using Eq.~(\ref{ai}), we obtain
\begin{align}
\label{a12}
\Gamma=\frac{2 \left({R_s}^2 \left[ 1+\rho_c \right] -\rho_c r^2 \right)}{{R_s}^2}\, .
\end{align}
In Figure~\ref{Fig:7}, we have depicted the adiabatic index $\Gamma$.
As it is clear from Figure~\ref{Fig:7}, the value of $\Gamma$ is greater than $\frac{4}{3}$
throughout the stellar interior and therefore the stability condition is satisfied.

\section{Discussion and conclusions}\label{S5}


In the present research, we considered the spherically symmetric and static configuration of the compact star by using the
{ Einstein-Gauss-Bonnet gravity coupled with a scalar field}.
In our formulation, for any given { spherically symmetric and static profile of the energy density $\rho$} and for arbitrary EoS of matter,
we can construct the model which reproduces the profile.
Because the profile of the energy density { determines} the mass $M$ and the radius $R_s$ of the compact star,
an arbitrary relation between the mass $M$ and the radius $R_s$ of the compact star can be realized by adjusting the potential $V(\xi)$
and the coefficient function $f(\xi)$ of the Gauss-Bonnet term in (\ref{g2}).
This could be regarded as { a} degeneracy between the EoS and the functions $V(\xi)$ and $f(\xi)$ characterizing the model,
which tells that only { the} mass-radius relation is insufficient to constrain the model.

As a concrete example, by using the polytrope EoS (\ref{polytrope}) and assuming the profile of the energy density $\rho(r)$ in (\ref{anz1}),
we have constructed a model and have discussed the property.
The derived analytic solution is scanned analytically and graphically by using different tests to monitor the physical { relevances} of the derived solution.

In this regard, we discover that the { energy density} and pressure decrease as radial coordinate approach the surface of the star Figure~(\ref{Fig:1}).
This indicates clearly that the center of the star is highly compact and the model under consideration is
valid for the region outside the center of the stellar.
Additionally, we have explained analytically and graphically { in Figure}~(\ref{Fig:5})
that all { the energy conditions} are verified throughout the interior of stellar configuration.
Due to Herrera (1992) \cite{HERRERA1992206}, any stable solution must yield a square of sound speed, $v^2$, to lie in the interval $v^2\in [0, 1]$.
In this model, we have shown that the speed of sound lies in the required { interval, which shows} that { the solution obtained in our model} is stable.
Also, the calculation of the adiabatic index of our model is in excellent agreement with { the stability condition} as shown in Figure~\ref{Fig:7} (right panel).
We have depicted the mass-radius relation as shown in Figure~\ref{Fig:4} (middle panel).
As this figure { shows, the mass} $M$ takes a positive value through the interior of the stellar.
Additionally, it is easy to prove that as $r\rightarrow 0$, we obtain $M\rightarrow 0$,
which ensures that $M$ is regular at the core of the star.
We also showed that the procedure used in this study can significantly enhance the mass, recent corroborating observations of some massive two-solar mass neutron stars.
Moreover, as Bhuchdahl (1959) \cite{PhysRev.116.1027} has shown that for static spherically symmetric isotropic matter content,
the ratio between the mass to the radius should be $\frac M {R}<\frac{4}{9}$.
In this study, the ratio $\frac M {R_s}=\frac{1}{4}$ (see Figure~\ref{Fig:4} (middle panel)) shows that the Bhuchdahl condition is satisfied.
The compactification $C=\frac M {R_s}$ has been depicted in Figure~\ref{Fig:4} (right panel), which shows that the compactness should be $0<C<0.55$.
In Figure~\ref{Fig:6} (right panel), we have shown that the profile of the surface redshift is less than $2$ as required for
an isotropic model without a cosmological constant.
It has been shown that the upper limit of surface redshift is $2$, which is in agreement with our stellar configuration \cite{PhysRev.116.1027,1984grra.book.....S,bohmer2007minimum}.

In the present study, we have assumed that a physical energy density is given by Eq.~\eqref{anz1},
as it has a finite value at the center of the start $\rho_c$ and it is finite at the surface of the star, which is consistent with realistic compact stars.
Also, the metric potentials of this construction are physical because they are singularity free as $r\to 0$ and have finite values at the surface of the star.
Additionally, the mass of the star in the model under consideration has a finite value at the center as well as at the surface of the star.
Moreover, the constructed model yields a consistent form of the GB term, the scalar field $\xi$, the potential $V(\xi)$, and the coefficient function $f(\xi)$ have finite value as $r\to 0$.
Also, we have shown that the model under consideration is stable and its adiabatic index is more than $\frac{4}{3}$, which is consistent with observation.

Remarkably, NICER observations of PSR J0030+0451 and PSR J0740+6020 offer indications against the more squeezable models.
The latter has significantly more mass than the former, although they are nearly the same size.
So it is reasonable to suppose some processes to rationalize the non-squeezability of a neutron star as its mass increases.
On the other hand, the presence of high-mass pulsars$\sim 2M_\odot$ such as PSR J0740+6020, is known to prefer violation
of the upper sound speed conformal limit $v^2\leq 1/3$ posing another challenge for theoretical models even in low-density cases,
as demonstrated by Bedaque and Steiner~\cite{Bedaque:2014sqa} (see also \cite{Cherman:2009tw,Landry:2020vaw}).
In their study, of the pulsar PSR J0740+6020, Legred et al.~\cite{Legred:2021hdx} concluded that the conformal sound speed is strongly violated
at the neutron star core, whereas $v^2= 0.75$ with density $ 3.60 {\rho}_\mathrm{nuc.}$.
It is important to mention that such an issue does not appear in our constructed model, as shown in Figure~\ref{Fig:4}~\subref{fig:dpr}. 

To conclude, as far as we know, that this is the first time to derive an analytic isotropic spherically symmetric interior solution in the frame of EGBS theory.
From the above analysis, we ensure that the derived solution in this study verified all
the physical requirements of any isotropic stellar configuration in the frame of this theory.
An isotropic model in the frame of Rastall's theory is derived using the technique of conformal killing vectors \cite{Abbas:2018mxs}.
In this model, the authors showed that the maximum value of the compactness in their model was $0.028742$ and the redshift was $0.09444$.
If we compare our results with the ones presented in \cite{Abbas:2018mxs}, we see that the compactness and redshift of our model are greater than
the ones presented in \cite{Abbas:2018mxs}.
This means that the non-linear form of the EoS has a greater effect on the structure of the model than the conformal killing vector.
An isotropic model is also constructed in the framework of $F(R,T)$, where $R$ is the Ricci scalar and $T$ is the trace of the energy-momentum tensor.
It was shown that the model constructed in \cite{Pappas:2022gtt} suffers from a violation of DEC, whereas it is satisfied in the model under consideration.
Moreover, it was shown that in the model constructed in \cite{Pappas:2022gtt}, its energy density configuration is non-uniform,
which corresponds to a quasi-constant density configuration, but our model did not possess a such defect. 

\appendix

\section{Explicit form of $f(r)$, $\xi(r)$, and $V(r)$}

In this Appendix, we give the explicit form and asymptotic forms of $f(r)$, $\xi(r)$, and $V(r)$.

The explicit form of $f(r)$ is given by
\begin{align}
\label{A1}
f(r)=&\, \frac{1}{16{R_s}^4}\int \left( \int \left[ \exp \left\{ -\int \left(2{r_0}^9K^2{\rho_c}^2{R_s}^2+2{r_0}^9{R_s}^2K\rho_c-M{R_s}^4r^6+3M^2{R_s}^{4 }r^5+10{\rho_c}^2r^{11}K^2
\right. \right. \right. \right. \nonumber \\
&\, +2{r_0}^6K^2{\rho_c}^2{R_s}^4M  +4 {r_0}^6K\rho_c {R_s}^4M+2 K\rho_c {R_s}^2r^9-47 {\rho_c}^2r^{10}K^2M+55 {\rho_c}^2r^9K^2M^2+30 {\rho_c}^2r^8K^2{r_0}^3 \nonumber \\
&\, +30 {\rho_c}^2r^5K^2{r_0}^6+10 {\rho_c}^2r^2K^2{r_0}^9 +2 K^2{\rho_c}^2{R_s}^2r^9-6 M^2{R_s}^4r^2{r_0}^3+M{R_s}^4r^3{r_0}^3-24 K\rho_c {R_s}^2Mr^5{r_0}^3 \nonumber \\
&\, +44 K\rho_c {R_s}^2M^2r^4{r_0}^3+M{R_s}^4K^2{\rho_c}^2r^3{r_0}^3 -6 M^2{R_s}^4K^2{\rho_c}^2r^2{r_0}^3+2 M{R_s}^4K\rho_c r^3{r_0}^3 \nonumber \\
&\, - 12 M^2{R_s}^4K\rho_c r^2{r_0}^3-24 K^2{\rho_c}^2{R_s}^2Mr^2{r_0}^6-24 K^2{\rho_c}^2{R_s}^2Mr^5{r_0}^3 +44 K^2{\rho_c}^2{R_s}^2M^2r^4{r_0}^3 \nonumber \\
&\, -24 K\rho_c {R_s}^2Mr^2{r_0}^6-M{R_s}^4K^2{\rho_c }^2r^6+3 M^2{R_s}^4K^2{\rho_c}^2r^5-2 M{R_s}^4K\rho_c r^6+6 M^2{R_s}^4K\rho_c r^5 \nonumber\\
&\, +6 K^2{\rho_c}^2{R_s}^2r^3{r_0}^6+6 K^2{\rho_c}^2{R_s}^2r^6{r_0}^3- 10 K^2{\rho_c}^2{R_s}^2M^2r^7+6 K\rho_c {R_s}^2r^3{r_0}^6+6 K\rho_c {R_s}^2r^6{r_0}^3 \nonumber \\
&\, \left. -10 K\rho_c {R_s}^2M^2r^7 -26 {\rho_c}^2r^4K^2M{r_0}^6-73 {\rho_c}^2r^7 K^2M{r_0}^3+10 {\rho_c}^2r^6K^2M^2{r_0}^3+2 {r_0}^6{R_s}^4M \right) \nonumber \\
&\, \times \left[ \left( M{R_s}^2 K\rho_c+M{R_s}^2+2 K\rho_c r^3 -5 K\rho_c M r^2+2 K\rho_c {r_0}^3 \right) \right. \left(2Mr^2 -r^3-{r_0}^3 \right) \left( r^3+{r_0}^3 \right) \nonumber \\
&\, \left. \left. \times \left( {R_s}^2+K\rho_c {R_s}^2-K\rho_c r^2 \right) r\right]^{-1}{dr}\right\} \left[6 {R_s}^8K^2{\rho_c}^3{r_0}^6
 -8 {R_s}^6{\rho_c}^2K^2{r_0}^6-8 {R_s}^6\rho_c K{r_0}^6+6 {R_s}^8{r_0}^6K{\rho_c}^2\right. \nonumber\\
&\, +2 {R_s}^8{r_0}^6K^3{\rho_c}^4+12 {R_s}^8r^3K{r_0}^3{\rho_c}^2+4 {R_s}^8r^3K^3{r_0}^3{\rho_c}^4
 -2 {R_s}^8r^3MK\rho_c-{R_s}^8r^3MK^2{\rho_c}^2 \nonumber\\
&\, + 12 {R_s}^8r^3K^2{r_0}^3{\rho_c}^3-16 r^9K^3{\rho_c}^4{R_s}^2{r_0}^3-12 r^9K^2{\rho_c}^3{R_s}^2{r_0}^3+36 r^7K^2{\rho_c}^3{R_s}^4{r_0}^3\nonumber\\
&\, +35 r^7K^2{\rho_c}^2{R_s}^4M+12 r^7K{\rho_c}^2{R_s}^4{r_0}^3+24 r^7K^3{\rho_c}^4{R_s}^4{r_0}^3-6 r^6K^2{\rho_c}^3{R_s}^2{r_0}^6-8 r^6K^3{\rho_c}^4{R_s}^2{r_0}^6 \nonumber\\
&\, -36 r^5{r_0}^3{R_s}^6{\rho_c}^3K^2-32 r^5K^2{\rho_c}^2{R_s}^4{r_0}^3+14 r^5{R_s}^6\rho_c MK+14 r^5{R_s}^6{\rho_c}^2K^2M-16 r^5{R_s}^6{\rho_c}^4K^3{r_0}^3 \nonumber \\
&\, -24 r^5{r_0}^3{R_s}^6{\rho_c}^2K +6 r^4K{\rho_c}^2{R_s}^4{r_0}^6+12 r^4K^3{\rho_c}^4{R_s}^4{r_0}^6+18 r^4K^2{\rho_c}^3{R_s}^4{r_0}^6-16 r^3{R_s}^6{\rho_c}^2K^2{r_0}^3 \nonumber \\
&\, -16 r^3{R_s}^6\rho_c K{r_0}^3-12 r^2{R_s}^6K{\rho_c}^2{r_0}^6 -18 r^2{R_s}^6K^2{\rho_c}^3{r_0}^6-8 r^2{R_s}^6 {\rho_c}^4K^3{r_0}^6-16 r^2{\rho_c}^2{R_s}^4K^2{r_0}^6 \nonumber \\
&\, -8 {R_s}^8K\rho_c M{r_0}^3-4 {R_s}^8K^2{\rho_c}^2M{r_0}^3+6 {R_s}^8r^6{\rho_c}^3K^2 +6 {R_s}^8r^6{\rho_c}^2K+2 {R_s}^8r^6{\rho_c}^4K^3+4 {R_s}^8r^3\rho_c {r_0}^3 \nonumber \\
&\, -6 r^{12}K^2{\rho_c}^3{R_s}^2-8 r^{12}K^3{\rho_c}^4{R_s}^2+6 r^{10}K{\rho_c}^2{R_s}^4+18 r^{10}K^2{\rho_c}^3{R_s}^4 +12 r^{10}K^3{\rho_c}^4{R_s}^4-12 r^8{R_s}^6{\rho_c}^2K \nonumber\\
&\, -18 r^8{R_s}^6{\rho_c}^3K^2 - 8 r^8{R_s}^6{\rho_c}^4K^3-16 r^8K^2{\rho_c}^2{R_s}^4-8 r^6{R_s}^6{\rho_c}^2K^2-8 r^6{R_s}^6\rho_c K +4 r^{11}K^3{\rho_c}^4{r_0}^3 \nonumber\\
&\, +2 r^8{K }^3{\rho_c}^4{r_0}^6-4 r^5{R_s}^6\rho_c \,{r_0}^3-2 r^2{R_s}^6{r_0}^6\rho_c+32\, r^2{R_s}^6K^2{\rho_c}^2{r_0}^3M+20\,r^{4 }K^2{\rho_c}^2{R_s}^4{r_0}^3M\nonumber\\
&\, \left. +32 r^2{R_s}^6K\rho_c {r_0}^3M-2 r^8{R_s}^6\rho_c+2 {R_s}^8r^6\rho_c-{R_s}^8r^3M+2 r^{14}K^3{\rho_c}^4+2 {R_s}^8\rho_c {r_0}^6-4 {R_s}^8{r_0}^3M \right] \nonumber \\
&\, \times \left\{ {R_s}^2+K\rho_c {R_s}^2 -K\rho_c r^2 \right\}^{-1} \left( 2Mr^2- r^3-{r_0}^3 \right)^{-1} \left( M{R_s}^2 K\rho_c+M{R_s}^2+2 K\rho_c r^3 -5 K\rho_c M r^2 \right. \nonumber \\
&\, \left. \left. \left. +2 K\rho_c {r_0}^3 \right)^{-1} \right]{dr}+16\,C \,{R_s}^4 \right)
\exp \left(\int \left\{2 {r_0}^9K^2{\rho_c}^2{R_s}^2+2 {r_0}^9{R_s}^2K\rho_c-M{R_s}^4r^6 +3 M^2{R_s}^4r^5 \right. \right. \nonumber \\
&\, +10 {\rho_c}^2r^{11}K^2+2 {r_0}^6K^2{\rho_c}^2{R_s}^4M+4 {r_0}^6K\rho_c {R_s}^4M +2 K\rho_c {R_s}^2r^9-47 {\rho_c}^2r^{10}K^2M + 55 {\rho_c}^2r^9K^2M^2 \nonumber\\
&\, +30 {\rho_c}^2r^8{ K}^2{r_0}^3+30 {\rho_c}^2r^5K^2{r_0}^6+ 10 {\rho_c}^2r^2K^2{r_0}^9+2 K^2{\rho_c }^2{R_s}^2r^9 -6 M^2{R_s}^4r^2{r_0}^3+M{R_s}^4r^3{r_0}^3 \nonumber\\
&\, -24 K\rho_c {R_s}^2Mr^5{r_0}^3+44 K\rho_c {R_s}^2M^2r^4{ r_0}^3+M{R_s}^4K^2{\rho_c}^2r^3{r_0}^ {3}-6 M^2{R_s}^4K^2{\rho_c}^2r^2{r_0}^3 +2 M{R_s}^4K\rho_c r^3{r_0}^3 \nonumber\\
&\, -12 M^2{R_s}^4K\rho_c r^2{r_0}^3-24 K^2{\rho_c}^2{R_s}^2Mr^2{r_0}^6-24 K^2{\rho_c}^2{R_s}^2Mr^5{r_0}^3+44 K^2{\rho_c}^2{{R_s}}^2M^2r^4{r_0}^3\nonumber\\
&\, -24 K\rho_c {R_s}^2Mr^2{r_0}^6-M{R_s}^4K^2{\rho_c}^2r^6+3 M^2{R_s}^4K^2{\rho_c}^2r^5-2 M{R_s}^4K{\rho_c} r^6+6 M^2{R_s}^4K\rho_c r^5 \nonumber\\
&\, +6 K^2{\rho_c}^2{R_s}^2r^3{r_0}^6+6 K^2{\rho_c}^2{R_s}^2r^6{r_0}^3-10 K^2{\rho_c}^2{R_s}^2M^2r^7+6 K\rho_c {R_s}^2 r^3{r_0}^6+6 K\rho_c {R_s}^2r^6{r_0}^3 \nonumber\\
&\, \left. -10 K\rho_c {R_s}^2M^2r^7-26 {\rho_c}^2 r^4K^2M{r_0}^6-73 {\rho_c}^2r^7K^2M{r_0}^3+10 {\rho_c}^2r^6K^2M^2{r_0}^3+2 {r_0}^6{R_s}^4M \right\} \nonumber \\
&\, \times \left[\left( M{R_s}^2K\rho_c+M{R_s}^2+2 K\rho_c r^3-5 K\rho_c Mr^2+2 K{\rho_c}\,{r_0}^3 \right) \left( 2Mr^2-r^3-{r_0}^3 \right) \right. \nonumber\\
&\, \left. \left. \times \left( r^3+{r_0}^3 \right) \left( {R_s}^2+K\rho_c {R_s}^2-K\rho_c r^2 \right) r\right]dr\right) dr +C_1\,.
\end{align}
The asymptotic form of $f(r)$ as $r\to 0$ takes the form,
\begin{align}
\label{A2}
f(r)\approx C_1+C_2r+C_3r^2+C_4r^4+C_5r^5\,,
\end{align}
where $C_2$, $\cdots$, $C_5$ are constants structured by $K$, $\rho_c$. and $r_0$.

The form of $\xi(r)$, after using the data given in Eqs.~(\ref{anz1}), (\ref{FRN3p1BC1}), and (\ref{matching}) takes the form:
\begin{align}
\label{A3}
\xi(r)=\, & \pm \frac{\sqrt{2}}{{R_s}^2}\int \left[\left( \left\{64 {R_s}^4C_2 K\rho_c r^{10}-48 {R_s}^6M^2r^2C_1 {r_0}^3-128 {R_s}^6M^2r^3{C_2}{r_0}^3+16 {R_s}^6MC_1 {r_0}^6
\right. \right. \right.\nonumber \\
&\, +8 {R_s}^6Mr^3C_1 {r_0}^3 +48 {R_s}^6Mr^4C_2 {r_0}^3+48 {R_s}^6MrC_2 {r_0}^6-128 {R_s}^6M^2r^3C_2 {r_0}^3K\rho_c-48 {R_s}^6M^2r^2C_1 {r_0}^3K\rho_c \nonumber\\
&\, +16 {R_s}^6MC_1 {r_0}^6K\rho_c +{R_s}^4[240M^2r^4C_1 {r_0}^3K\rho_c-176Mr^2C_1{r_0}^6K\rho_c-688Mr^6C_2{r_0}^3K\rho_c \nonumber\\
&\, \left. -368M r^3C_2{r_0}^6K\rho_c+48{R_s}^2MrC_2 {r_0}^6K\rho_c \right] +{R_s}^4 \left[48 {R_s}^2Mr^4C_2 {r_0}^3K\rho_c+512M^2r^5C_2 {r_0}^3K\rho_c \right. \nonumber\\
&\, \left. +8 {R_s}^2Mr^3C_1 {r_0}^3K \rho_c-328Mr^5C_1 {r_0}^3K\rho_c+32K\rho_c r^9C_1 \right] -K^2{\rho_c}^3r^{16}+Mr^7{R_s}^6+\rho_c r^{10}{R_s}^6 \nonumber \\
&\, -3^2{\rho_c}^3r^{12}{R_s}^4+K^2{\rho_c }^3r^{10}{R_s}^6+3 K^2{\rho_c}^3r^{14}{R_s}^2-3 K^2{\rho_c}^3r^{13}{r_0}^3
 - 3 K^2{\rho_c}^3r^{10}{r_0}^6-K^2{\rho_c}^3r^7{r_0}^9 \nonumber\\
&\, +K^2{\rho_c}^3r{r_0}^9{R_s}^6-9 { K}^2{\rho_c}^3r^9{R_s}^4{r_0}^3-9 K^2{\rho_c}^3r^6{R_s}^4{r_0}^6-3 K^2{\rho_c}^3r^3{R_s}^4{r_0}^9 +3 K^2{\rho_c}^3r^7{R_s}^6{r_0}^3 \nonumber\\
&\, +3 K^2{\rho_c}^3r^4{R_s}^6{r_0}^6+9 K^2{\rho_c}^3r^{11}{R_s}^2{r_0}^3+9 K^2{\rho_c}^3r^8{R_s}^2{{ r_0}}^6+3 K^2{\rho_c}^3r^5{R_s}^2{r_0}^9+Mr^7{R_s}^6\rho_c K \nonumber\\
&\, +7 Mr^9{R_s}^4K\rho_c+6 {\rho_c}^2r^7{R_s}^6K{r_0}^3+6 {\rho_c}^2r^4{R_s}^6K{r_0}^6+2 {\rho_c}^2r{R_s}^6K{r_0}^9-12 {\rho_c}^2r^9{R_s}^4K{r_0}^3 \nonumber\\
&\, -12 {\rho_c}^2r^6{R_s}^4K{r_0}^6-4 {\rho_c}^2r^3{R_s}^4K{r_0}^9+6 {\rho_c}^2r^{11}{R_s}^2K{r_0}^3+6 {\rho_c}^2r^8{R_s}^2K{r_0}^6+2 {\rho_c}^2r^5{R_s}^2K{r_0}^9 \nonumber\\
&\, -12 {R_s}^4K\rho_c r^7{r_0}^3-12 {R_s}^4K\rho_c r^4{r_0}^6+{R_s}^4 \left[10Mr^3{r_0}^6K\rho_c-4K\rho_c r{r_0}^9-Mr^4\rho_c \,K{r_0}^3+17\,Mr^6{r_0}^3K\rho_c \right. \nonumber \\
&\, \left. - 2 Mr{R_s}^2\rho_c K{r_0}^6+96K\rho_c r^6C_1{r_0}^3 \right]-152 {R_s}^4Mr^8C_1 K\rho_c+368 {R_s}^4M^2r^8C_2 K\rho_c+24 {R_s}^6M^2r^5C_1 K\rho_c \nonumber\\
&\, -320 {R_s}^4Mr^9C_2 K\rho_c+16 {R_s}^6M^2r^6C_2 K \rho_c+168 {R_s}^4M^2r^7C_1 K\rho_c-8 {R_s}^6Mr^6C_1 K\rho_c+192 {R_s}^4C_2 K\rho_c r^7{r_0}^3 \nonumber \\
&\, +64 {R_s}^4C_2 K\rho_c r{r_0}^9+192 {R_s}^4C_2 K\rho_c  r^4{r_0}^6 +32 {R_s}^4K\rho_c C_1 {r_0}^9+96 {R_s}^4K\rho_c r^3C_1 {r_0}^6- Mr^4{R_s}^6{r_0}^3 \nonumber\\
&\, -2 Mr{R_s}^6{r_0}^6 +3 \rho_c r^7{R_s}^6{r_0}^3+ 3\rho_c r^4{R_s}^6{r_0}^6+\rho_c r{R_s}^6{r_0}^9 +2 {\rho_c}^2r^{10}{R_s}^6K-4 {\rho_c}^2r^{12}{R_s}^4K \nonumber\\
&\, -3 \rho_c r^9{R_s}^4{r_0}^{3 }-3\,\rho_c r^6{R_s}^4{r_0}^6-\rho_c r^3{R_s}^4{r_0}^9+2 {\rho_c}^2r^{14}{R_s}^2K-4 {R_s}^4K\rho_c r^{10} -\rho_c r^{12}{R_s}^4 \nonumber\\
&\, \left. \left. -8 {R_s}^6Mr^6C_1+16 {R_s}^6M^2r^6C_2+24 {R_s}^6M^2r^5C_1\right\} \right) \left\{\sqrt{r} \sqrt {-r^3+2 Mr^2-{r_0}^3} \right. \nonumber\\
&\, \left. \left. \times \sqrt {{R_s}^2+K\rho_c {R_s}^2-K\rho_c r^2} \left( r^2-rr_0+{r_0}^2 \right) \left( r+r_0 \right) \right\}^{-1} \right]{dr}+{C_6 }\,,
\end{align}
The asymptotic form of $\xi(r)$ as $r\to 0$ takes the form:
\begin{align}
\label{A4}
\xi(r) \approx C_6+C_7r^2+C_8 r^3\,,
\end{align}3K
where $C_6$, $C_7$, and $C_8$ are structured by the constants $K$, $R_s$, $\rho_c$, and $r_0$.

Finally, we calculate the explicit form of $V(r)$ after using the data given in Eqs (\ref{FRN3p1BC1}), (\ref{matching}), (\ref{anz1}) and obtain,
\begin{align}
\label{A5}
V(r)=&\,-\left\{8 {R_s}^6Mr^3C_1 {r_0}^3-64 {R_s}^4C_2 K\rho_c r^{10}-48 {R_s}^6M^2r^2C_1 {r_0}^3-128 {R_s}^6M^2r^3C_2 {r_0}^3 +16 {R_s}^6MC_1 {r_0}^6 \right. \nonumber\\
&\, +48 {R_s}^6Mr^4C_2 {r_0}^3+48 {R_s}^6MrC_2 {r_0}^6-128 {R_s}^6M^2r^3C_2 {r_0}^3K\rho_c-48 {R_s}^6M^2r^2C_1 {r_0}^3K\rho_c \nonumber\\
&\, +16 {R_s}^6MC_1 {r_0}^6K\rho_c + 144 {R_s}^4Mr^2C_1 {r_0}^6K\rho_c+592 {R_s}^4Mr^6C_2 {r_0}^3K\rho_c+272 {R_s}^4Mr^3C_2 {r_0}^6K\rho_c \nonumber\\
&\, +48 {R_s}^6MrC_2 {r_0}^6K\rho_c -144 {R_s}^4M^2r^4C_1 {r_0}^3K\rho_c+48 {R_s}^6Mr^4C_2 {r_0}^3K\rho_c-256 {R_s}^4M^2r^5C_2 {r_0}^3K\rho_c \nonumber\\
&\, +8 {R_s}^6Mr^3C_1 {r_0}^3K\rho_c +312 {R_s}^4Mr^5C_1 {r_0}^{3 }K\rho_c-32 {R_s}^4K\rho_c r^9C_1+K^2{\rho_c}^3r^{16}-Mr^7{R_s}^6+\rho_c r^{10}{R_s}^6 \nonumber\\
&\, +3 K^2{\rho_c}^3r^{12}{R_s}^4-K^2{\rho_c}^3r^{10}{R_s}^6 -3 K^2{\rho_c}^3r^{14}{R_s}^2+3 K^2{\rho_c}^3r^{13}{r_0}^3+3 K^2{\rho_c}^3r^{10}{r_0}^6+K^2{\rho_c}^3 r^7{r_0}^9 \nonumber\\
&\, -K^2{\rho_c}^3r{r_0}^9{R_s}^6+9 K^2{\rho_c}^3r^9{R_s}^4{r_0}^3 +9 K^2{\rho_c}^3r^6{R_s}^4{r_0}^6+3 K^2{\rho_c}^3r^3{R_s}^4{r_0}^9 -3 K^2{\rho_c}^3r^7{R_s}^6{r_0}^3 \nonumber\\
&\, -3 K^2{\rho_c}^3r^4{R_s}^6{r_0}^6-9 K^2{\rho_c}^3r^{11}{R_s}^2{r_0}^3 -9 K^2{\rho_c}^3r^8{R_s}^2{r_0}^6-3 K^2{\rho_c}^3r^5{R_s}^2{r_0}^9-Mr^7{R_s}^6\rho_c K \nonumber\\
&\, -7 Mr^9{R_s}^4K\rho_c+12 {R_s}^4K\rho_c r^7{r_0}^3+ 12 {R_s}^4K\rho_c r^4{r_0}^6+4 {R_s}^4 K\rho_c r{r_0}^9-5 Mr^4{R_s}^6\rho_c K{r_0}^3 \nonumber\\
&\, -11 Mr^6{R_s}^4{r_0}^3K\rho_c-4 Mr {R_s}^6\rho_c K{r_0}^6-4 Mr^3{R_s}^4 {r_0}^6K\rho_c -96 {R_s}^4K\rho_c r^6C_1 {r_0}^3+168 {R_s}^4Mr^8C_1 K\rho_c \nonumber\\
&\, -400 {R_s}^4M^2r^8C_2 K\rho_c+24 {R_s}^6 M^2r^5C_1 K\rho_c+320 {R_s}^4Mr^9C_2 K\rho_c +16 {R_s}^6M^2r^6C_2 K\rho_c \nonumber\\
&\, - 216 {R_s}^4M^2r^7C_1 K\rho_c-8 {R_s}^6Mr^6C_1 K\rho_c -192 {R_s}^4C_2 K\rho_c r^7{r_0}^3-64 {R_s}^4C_2 K\rho_c r{r_0}^9 \nonumber \\
&\, -192 {R_s}^4C_2 K\rho_c r^4{r_0}^6 -32 {R_s}^4K\rho_c C_1 {r_0}^9- 96 {R_s}^4K\rho_c r^3C_1 {r_0}^6-5 Mr^4{R_s}^6{r_0}^3-4 Mr{R_s}^6{r_0}^6 \nonumber\\
&\, +3\rho_c r^7{R_s}^6{r_0}^3+3 \rho_c r^4 {R_s}^6{r_0}^6+\rho_c r{R_s}^6{r_0}^9 -3 \rho_c r^9{R_s}^4{r_0}^3-3 \rho_c r^6{R_s}^4{r_0}^6 -\rho_c r^3{R_s}^4{r_0}^9 \nonumber\\
&\, \left. +4 {R_s}^4K\rho_c r^{10}-\rho_c r^{12}{R_s}^4-8 {R_s}^6Mr^6C_1+16 {R_s}^6 M^2r^6C_2+24 {R_s}^6M^2r^5C_1 \right\} \nonumber \\
&\, \times \left\{{R_s}^4r \left( {R_s}^2+K\rho_c {R_s}^2-K\rho_c r^2 \right) \left( r^3+{r_0}^3 \right)^3\right\}^2\,.
\end{align}
The asymptotic form of $V(r)$ as $r\to 0$ takes the form:
\begin{align}
\label{A6}
V(r)\approx C_9+C_{10}r+C_{11}r^2\,,
\end{align}
where $C_9$, $C_{10}$, and $C_{11}$ are structured by the constants $K$, $R_s$, $\rho_c$, and $r_0$.

%

\end{document}